\documentclass[preprint]{aastex62}
\usepackage{float}

\usepackage{appendix}
\usepackage{mathtools}
\usepackage{epsfig}
\usepackage{array}
\usepackage{graphicx}
\usepackage{graphics}
\usepackage{subfigure}
\usepackage[latin1]{inputenc}
\usepackage{latexsym}
\newcommand{\nd}{\multicolumn{1}{c}{$\dots$}}

\shorttitle{SN~Ia 2014J} \shortauthors{Li et al.}

\def\gsim{\;\lower4pt\hbox{${\buildrel\displaystyle >\over\sim}$}\;}
\def\lsim{\;\lower4pt\hbox{${\buildrel\displaystyle <\over\sim}$}\;}
\def\grls{\;\lower4pt\hbox{${\buildrel\displaystyle >\over <}$}\;}

\begin{document}

\title{Observations of Type Ia Supernova 2014J for Nearly 900 Days and Constraints on Its Progenitor System}
\correspondingauthor{Xiaofeng Wang}
\email{wang\_xf@mail.tsinghua.edu.cn}

\author{Wenxiong Li}
\affil{Physics Department and Tsinghua Center for Astrophysics (THCA), Tsinghua University, Beijing, 100084, China}

\author{Xiaofeng Wang}
\affil{Physics Department and Tsinghua Center for Astrophysics (THCA), Tsinghua University, Beijing, 100084, China}
\author{Maokai Hu}
\affil{Purple Mountain Observatory, Chinese Academy of Sciences, Nanjing 210034, China}
\author{Yi Yang}
\affil{Department of Particle Physics and Astrophysics, Weizmann Institute
of Science, Rehovot 76100, Israel}
\author{Jujia Zhang}
\affil{Yunnan Astronomical Observatory of China, Chinese Academy of Sciences, Kunming, 650011, China}
\affil{Key Laboratory for the Structure and Evolution of Celestial Objects, Chinese Academy of Sciences, Kunming 650011, China}
\author{Jun Mo}
\affil{Physics Department and Tsinghua Center for Astrophysics (THCA), Tsinghua University, Beijing, 100084, China}
\author{Zhihao Chen}
\affil{Physics Department and Tsinghua Center for Astrophysics (THCA), Tsinghua University, Beijing, 100084, China}
\author{Tianmeng Zhang}
\affil{National Astronomical Observatory of China, Chinese Academy of Sciences, Beijing, 100012, China}
\author{Stefano Benetti}
\affil{INAF - Osservatorio Astronomico di Padova, Vicolo dell'Osservatorio 5, 35122 Padova, Italy}
\author{Enrico Cappellaro}
\affil{INAF - Osservatorio Astronomico di Padova, Vicolo dell'Osservatorio 5, 35122 Padova, Italy}
\author{Nancy Elias-Rosa}
\affil{Institute of Space Sciences (ICE, CSIC), Campus UAB, Carrer de Can Magrans s/n, 08193 Barcelona, Spain}
\affil{Institut d'Estudis Espacials de Catalunya (IEEC), c/Gran Capit\'a 2-4, Edif. Nexus 201, 08034 Barcelona, Spain
}
\author{Jordi Isern}
\affil{Institute of Space Sciences (ICE, CSIC), Campus UAB, Carrer de Can Magrans s/n, 08193 Barcelona, Spain}
\affil{Institut d'Estudis Espacials de Catalunya (IEEC), c/Gran Capit\'a 2-4, Edif. Nexus 201, 08034 Barcelona, Spain
}
\author{Antonia Morales-Garoffolo}
\affil{Department of Applied Physics, University of C\'adiz, Campus of Puerto Real, E-11510, C\'adiz, Spain}
\author{Fang Huang}
\affil{Department of Astronomy, School of Physics and Astronomy,  Shanghai Jiao Tong University, Shanghai 200240, China}
\affil{Key Laboratory for the Structure and Evolution of Celestial Objects, Chinese Academy of Sciences, Kunming 650011, China}

\author{Paolo Ochner}
\affil{Dipartimento di Fisica e Astronomia G. Galilei, Universita' di Padova, Vicolo dell'Osservaotio 3,35122, Padova, Italy}
\author{Andrea Pastorello}
\affil{INAF - Osservatorio Astronomico di Padova, Vicolo dell'Osservatorio 5, 35122 Padova, Italy}
\author{Andrea Reguitti}
\affil{Departamento de Ciencias Fisicas, Universidad Andr\'es Bello, Fernandez Concha, 700, Santiago, Chile}
\author{Leonardo Tartaglia}
\affil{Department of Astronomy, Stockholm University, SE 10691 Stockholm, Sweden}
\author{Giacomo Terreran}
\affil{Center for Interdisciplinary Exploration and Research in Astrophysics (CIERA) and Department of Physics and Astronomy, Northwestern University, Evanston, IL 60208, USA}
\author{Lina Tomasella}
\affil{INAF - Osservatorio Astronomico di Padova, Vicolo dell'Osservatorio 5, 35122 Padova, Italy}

\author{Lifan Wang}
\affil{Purple Mountain Observatory, Chinese Academy of Sciences, Nanjing 210034, China}
\affil{George P. and Cynthia Woods Mitchell Institute for Fundamental Physics $\&$ Astronomy, Texas A. $\&$ M. University, Department of Physics and Astronomy, 4242 TAMU, College Station, TX 77843, USA}


\begin{abstract}
We present extensive ground-based and $Hubble~Space~Telescope$ ($HST$)  photometry of the highly reddened, very nearby type Ia supernova (SN Ia) 2014J in M82, covering the phases from 9 days before to about 900 days after the $B$-band maximum. SN 2014J is similar to other normal SNe Ia near the maximum light, but it shows flux excess in the $B$ band in the early nebular phase. This excess flux emission can be due to light scattering by some structures of circumstellar materials located at a few 10$^{17}$ cm, consistent with a single degenerate progenitor system  or a double degenerate progenitor system with mass outflows in the final evolution or magnetically driven winds around the binary system. At t$\sim$+300 to $\sim$+500 days past the $B$-band maximum, the light curve of SN 2014J shows a faster decline relative to the $^{56}$Ni decay. Such a feature can be attributed to the significant weakening of the emission features around [Fe III] $\lambda$4700 and [Fe II] $\lambda$5200 rather than the positron escape as previously suggested. Analysis of the $HST$ images taken at t$>$600 days confirms that the luminosity of SN 2014J maintains a flat evolution at the very late phase. Fitting the late-time pseudo-bolometric light curve with radioactive decay of  $^{56}$Ni, $^{57}$Ni and $^{55}$Fe isotopes, we obtain the mass ratio $^{57}$Ni/$^{56}$Ni as $0.035 \pm 0.011$, which is consistent with the corresponding value predicted from the 2D and 3D delayed-detonation models. Combined with early-time analysis, we propose that delayed-detonation through single degenerate scenario is most likely favored for SN 2014J. 

\end{abstract}

\section{Introduction}
Type Ia supernovae (SNe Ia) are important tools to measure cosmological expansion \citep{1998AJ....116.1009R}. The progenitors of SNe Ia are believed to arise from a white dwarf (WD) inhabiting a binary system, with a mass close to the Chandrasekhar limit \citep{1960ApJ...132..565H}. However, the explosion mechanisms and binary evolution scenarios for SNe Ia are not well understood \citep{2011NatCo...2E.350H,2014ARA&A..52..107M}. The two common scenarios include (i) double-degenerate (DD) with one white dwarf tidally disrupting the WD companion and accreting its material \citep{1984ApJS...54..335I,1984ApJ...277..355W}, and (ii) single-degenerate (SD) where the WD accretes matter from a main sequence star \citep{1992A&A...262...97V}, a subgiant \citep{2004MNRAS.350.1301H}, a helium star \citep{2009MNRAS.395..847W,2013A&A...554A..54G} or a red giant companion \citep{1973ApJ...186.1007W,1982ApJ...253..798N,2011A&A...530A..63P}. One prediction of the SD scenario is that considerable amount of circumstellar material (CSM) should be accumulated around the progenitor system via stellar wind of a companion star or successive nova eruptions, although the DD and core-degenerate models are also argued to be able to form nearby CSM \citep{2015MNRAS.447.2803L,2015MNRAS.450.1333S}.

The CSM formed by outflowing materials would result in blueshifted and evolving narrow interstellar absorption features such as the Na I doublet. Variable Na~I D absorption was initially reported by  \cite{2007Sci...317..924P} in the spectra of SN 2006X. After that, several SNe Ia also showed variations in the Na I D absorption, including SN 2007le \citep{2009ApJ...702.1157S} and 
PTF 11kx \citep{2012Sci...337..942D}. There are also some statistical studies showing blueshifted narrow absorption features of Na I D in the high-resolution spectra of some SNe Ia \citep{2011Sci...333..856S, 2013MNRAS.436..222M,2014MNRAS.443.1849S}, which are associated with the subclass characterized by higher photospheric velocities \citep{2009ApJ...699L.139W,2013Sci...340..170W}. Both results indicate that some SNe Ia are surrounded by CSM, favoring their SD progenitor system origin \citep{2012ApJ...752..101F,2017MNRAS.471..491H}. 

Another effect due to the presence of surrounding CSM is the extinction along the line of sight and the scattering of the SN light. The CSM can scatter photons back to the  line of sight  and hence reduce the total extinction and $R_V$ \citep{2005ApJ...635L..33W, 2008ApJ...686L.103G}. Based on analysis of a large sample of Na I doublet absorption features in SN Ia spectra and light curve evolution in the early nebular phase, \citet{2018arXiv181011936W} provide evidence that SNe Ia with higher photospheric velocity (HV) likely have CS dust at a distance of about 2$\times$10$^{17}$ cm, implying that this subclass may have a single-degenerate origin.

Very early observations of SNe Ia are promising ways to constrain their progenitor systems. Different scenarios discuss the diversity of the early evolution of SNe Ia \citep{2010ApJ...708.1025K, 2016ApJ...826...96P,2014ApJ...794...37M,2017Natur.550...80J}. A growing number of SNe Ia with early phase observations have been studied \citep{2013ApJ...778L..15Z,2015Natur.521..328C,2016ApJ...820...92M,2017ApJ...845L..11H,2018ApJ...852..100M}. SN 2018oh is the only spectroscopically-confirmed normal SN Ia with $Kepler$ Space Telescope high-cadence photometry since explosion \citep{2019ApJ...870...12L}. Detailed studies reveal there is a bump in the early light curve of SN 2018oh, it may stem from non-degenerate companion interaction \citep{2019ApJ...870L...1D} or $^{56}$Ni radiation from the outer part of the ejecta \citep{2019ApJ...870...13S}.

The other way to distinguish the SD and DD scenario is to observe late-time evolution of SN Ia. The early-time light curve of SN Ia is powered by the radioactive decay chain $^{56}$Ni $\rightarrow$ $^{56}$Co $\rightarrow$ $^{56}$Fe with half-life $t_{1/2} \sim$ 6 and 77 days, respectively. 
With the expansion of the ejecta, the column density decreases as $t^{-2}$. Therefore, after t$\sim$200 days, the ejecta are almost transparent to $\gamma$-rays, and positron emission begins to dominate the heating process \citep{1979ApJ...230L..37A,1999ApJS..124..503M}. Assuming complete positron trapping, the light curve at this stage should follow the decay rate of $^{56}$Co, i.e., 0.98 mag per 100 days.

At very late phase, the light curves of some SNe Ia are found to flatten compared to the $^{56}$Co decay, including SN 1992A \citep{1997A&A...328..203C}, SN 2003hv \citep{2009A&A...505..265L}, SN 2011fe \citep{2014ApJ...796L..26K}, SN 2012cg \citep{2016ApJ...819...31G}, ASASSN-14lp \citep{2018ApJ...866...10G}, and SN 2015F \citep{2018ApJ...859...79G}. \cite{2009MNRAS.400..531S} suggested that the additional energy should come from the long-lived decay chains $^{57}$Co $\rightarrow$ $^{57}$Fe ($t_{1/2} \sim$ 272 days) or $^{55}$Fe $\rightarrow$ $^{55}$Mn ($t_{1/2} \sim$ 1000 days). The ratio of these isotopes relative to $^{56}$Ni depends on the density of the progenitor which is different in SD and DD models \citep{2012ApJ...750L..19R}. Higher $^{57}$Ni/$^{56}$Ni mass ratios are predicted by WD models of higher central density. This is because $^{57}$Ni is a neutron-rich isotope which has higher abundance in neutron rich environments such as near-Chandrasekhar-mass, delayed-detonation explosion models \citep{1989MNRAS.239..785K,2013MNRAS.429.1156S} as expected for the single degenerate channel. Therefore, these ratios can provide clues to the progenitor models. 

There are also other explanations for the flattening of the light curves. For example, an unresolved light echo, due to scattering of the SN light by nearby CSM or interstellar dust, can naturally increase the luminosity. \cite{1993ApJ...408L..25F} suggested that the recombination and cooling time scale is longer than radioactive decay as a result of decreasing density of the ejecta, the so-called ``freeze-out" effect, which can make the light curves flat at late time. Combining this  ``freeze-out" effect  with the $^{56}$Co decay, \cite{2015ApJ...814L...2F} and \cite{2017MNRAS.472.2534K} provided reasonable explanations for the very late-time spectrum and light curve of SN 2011fe. 

SN 2014J is a very nearby SN Ia, the distance being $\sim$3.53 Mpc \citep{2009ApJS..183...67D}. This provides an excellent opportunity to study its progenitor through the detection of dusty environment and the analysis of late-time light curve evolution. This bright SN has been followed up by many instruments, covering $\gamma$-rays by \textit{INTEGRAL} \citep{2014Natur.512..406C,2015ApJ...812...62C,2014Sci...345.1162D,2015A&A...574A..72D,2016A&A...588A..67I} and \textit{Suzaku} \citep{2016ApJ...823...43T}, X-rays \citep{2014ApJ...790...52M}, UV \citep{2015ApJ...805...74B,2014MNRAS.443.2887F}, optical \citep{2015ApJ...799..105S,2015ApJ...799..106G,2016A&A...585A..19B}, near-infrared (NIR) \citep{2015ApJ...798...39M,2015ApJ...804...66V,2016ApJ...822L..16S}, mid-infrared (MIR) \citep{2017MNRAS.466.3442J,2015ApJ...798...93T}, and radio bands \citep{2014ApJ...792...38P}, polarimetric observations have also been presented  \citep{2014ApJ...795L...4K,2015A&A...577A..53P,2016ApJ...828...24P,2018ApJ...854...55Y}.

The heavy reddening towards SN 2014J has inspired plentiful studies focusing on its dusty environment. With the UV and optical light curves and spectra obtained with \textit{Swift}, \citet{2015ApJ...805...74B} concluded that the large reddening mainly originates from the absorption features of the interstellar medium (ISM), which is also confirmed by the light echo emerging after $\sim$200 days \citep{2015ApJ...804L..37C, 2017ApJ...834...60Y}. By measuring the continuum polarization of SN 2014J, \cite{2015A&A...577A..53P} came to a similar conclusion. \cite{2014ApJ...788L..21A} found that the unusual reddening behavior seen in SN 2014J can be explained by non-standard ISM dust with $R_V = 1.4 \pm 0.1$ or by power-law extinction of CSM \citep{2005ApJ...635L..33W}. \cite{2014MNRAS.443.2887F} used a hybrid model including both ISM and CSM to explain the extinction curve. \cite{2015ApJ...801..136G} identified time-varying potassium lines and attributed them to the CSM origin. Based on the $Spitzer$ MIR data, \cite{2017MNRAS.466.3442J} placed an upper limit on the pre-existing circumstellar dust, with M$_{CSM}\le 10^{-5}$ M$_\odot$ and a distance of  $r\sim 10^{17}$ cm from the SN. By studying the late-time polarimetry of SN 2014J with the $Hubble~Space~Telescope~(HST)$ ACS/WFC observations, \cite{2018ApJ...854...55Y} concluded that at least $\sim 10^{-6} ~\rm M\odot$ of circumstellar dust is located at a distance of 5$\times$10$^{17}$ cm from SN 2014J. 

\cite{2018ApJ...852...89Y} also analyzed the late-time $HST$ photometry and found that both the $F665W$-band and the bolometric light curve of SN 2014J exhibited flattening behavior. Recently, \cite{2018arXiv181007258G} confirms the late-time flattening in $F555W$ and $F438W$ bands of \textit{HST} Wide Field Camera 3 UVIS channel (WFC3/UVIS)
.

In this paper, we present extensive photometry of SN 2014J from ground-based telescopes and $HST$, and analyze the light curve evolution by comparison with other well-observed SNe Ia. Photometric observations are addressed in Section \ref{observation}. In Section \ref{early}, we examine the light curves near the maximum light and in the early nebular phase. In Section \ref{late}, we examine the very late-time evolution and explore its possible origins. We discuss and conclude in Section \ref{conclusion}.

%

\section{OBSERVATIONS}\label{observation}

\subsection{Ground-based Photometry}
Our ground-based optical photometry of SN 2014J was obtained with the following telescopes: (1) the 0.8~m THCA-NAOC Telescope \citep[TNT;][]{2012RAA....12.1585H} at Beijing Xinglong Observatory (BAO) in China; (2) the 2.4~m Lijiang Telescope (LJT) of Yunnan Astronomical Observatory (YNAO); (3) the 2.56~m Nordic Optical Telescope (NOT) of Roque de los Muchachos Observatory in the Canary Islands; (4) the 0.8~m Telescopi Joan Or{\'o} (TJO) locates at the Montsec Astronomical Observatory;  (5) the Asiago Copernico 1.82~m telescope (COP) with AFOSC; (6) the Asiago Schmidt 67/92 (SCH) with SBIG and (7) the 3.58~m Telescopio Nazionale Galileo (TNG) with LRS. All CCD images were corrected for bias and flat field. Template subtraction has been applied to improve photometry to all images. The latest ground-based image with signals was taken with the Asiago Copernico 1.82~m telescope (COP)  at ~434 days relative to the B-band maximum light. And the template image used for galaxy subtraction was obtained on Jun. 19th 2017, corresponding to about 1234.0 days after the maximum light. This late-time COP V-band image with galaxy subtraction is shown in Figure 6(b),  together with the F555 $HST$ images. The SN was imaged in the \textit{UBVRI} bands with TNT, LJT, TJO, TNG and COP, \textit{BVRI} bands with NOT and SCH.


\begin{figure}[htbp]
\center
\includegraphics[width=\linewidth]{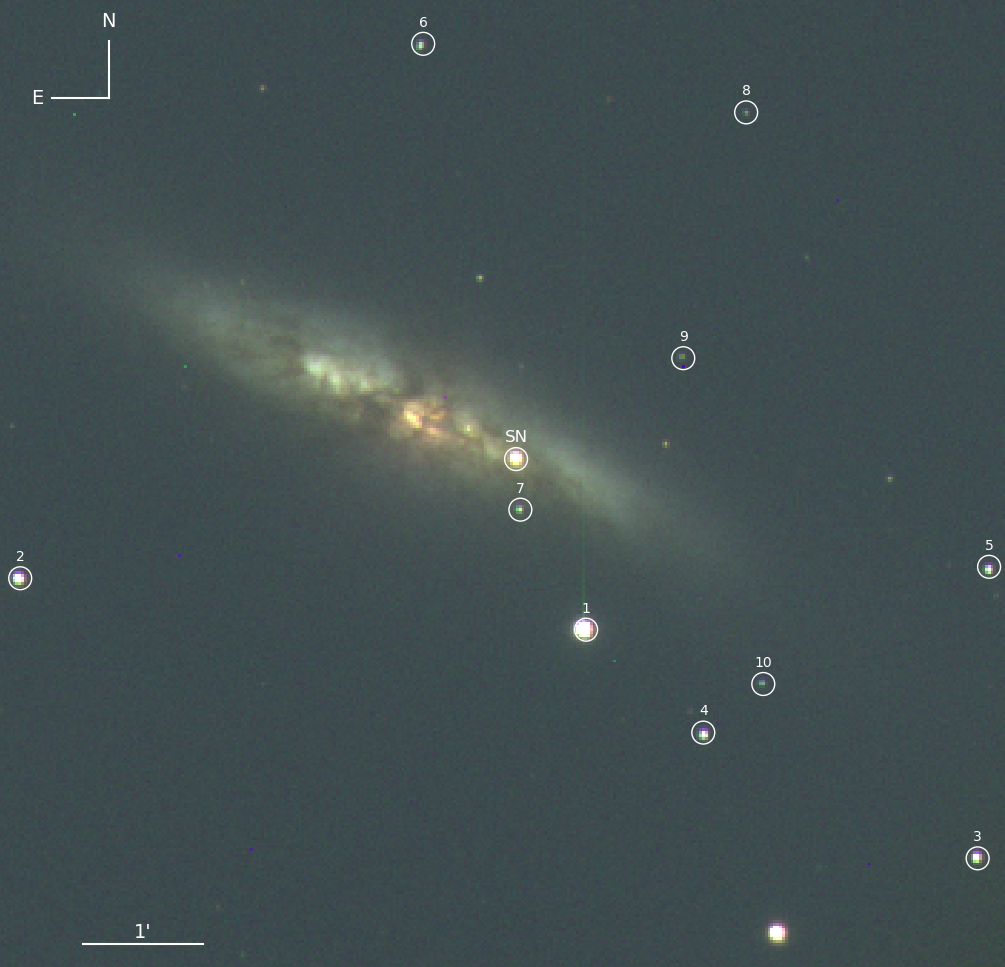}
\vspace{0.2cm}
\caption{SN 2014J in M82. The composite image was produced by combining $B$(blue)-, $V$ (green)- and $R$ (red)-band images obtained with LJT on 2014-01-22.8 UT. The reference stars listed in Table \ref{std} are numbered. North is up and east is to the left.}
\label{BVR} \vspace{-0.0cm}
\end{figure}

The photometry data were analyzed with the open source online photometry and astrometry codes SWARP \citep{2002ASPC..281..228B}, SCAMP \citep{2006ASPC..351..112B}, and SExtractor \citep{1996A&AS..117..393B}. We performed aperture photometry on the template subtracted SN images with SExtractor. The SN instrumental magnitudes were calibrated using Landolt standards stars. The photometric uncertainties including the Poisson noise of the signal and the photon noise of the local background. The field of SN 2014J taken with LJT is displayed in Figure \ref{BVR},  the latest template-subtracted image taken by COP on +434 days in $V$ band is displayed in Figure \ref{f555}(b),  and the final flux-calibrated magnitudes are listed in Table \ref{gphoto}.

\subsection{$HST$ Photometry}
Very late-time photometry of SN 2014J can be extracted from the archival images obtained through the $HST$ WFC3/UVIS programs (Proposal 13626, PI: Lawrence, 14146, PI: Lawrence and 14700, PI: Sugerman). The wavelength coverage of $F438W$ and $F555W$ filters is similar to that of \textit{B} and \textit{V} band, respectively, as shown in Figure \ref{BV_H}. The information on $HST$ observations is given in Table \ref{hst}. 

When multiple exposures were available at similar phases in the same filter, we combined them (using SCAMP and SWARP). As for the ground-telescope images, we performed aperture photometry with SExtractor. For each image, we chose a 6-pixel aperture for the photometry which does not include the flux from the light echo. However, for the last-epoch image obtained in $HST$ $F438W$ filter at +881 days from the maximum light (when the SN had faded enough), we adjusted the aperture size to 4 pixels to avoid background contamination. We applied aperture corrections according to \citet{2017wfc..rept...14D} to calculate the total flux of the SN. The photometric error added in quadrature includes uncertainties in Poisson noise, background noise and aperture correction. The resulting photometric analysis is presented in Table \ref{hphoto}.

\section{Evolution during the First 5 Months}\label{early}
\subsection{Light/Colour Curves}
\begin{figure}[htbp]
\center
\includegraphics[width=0.8\linewidth]{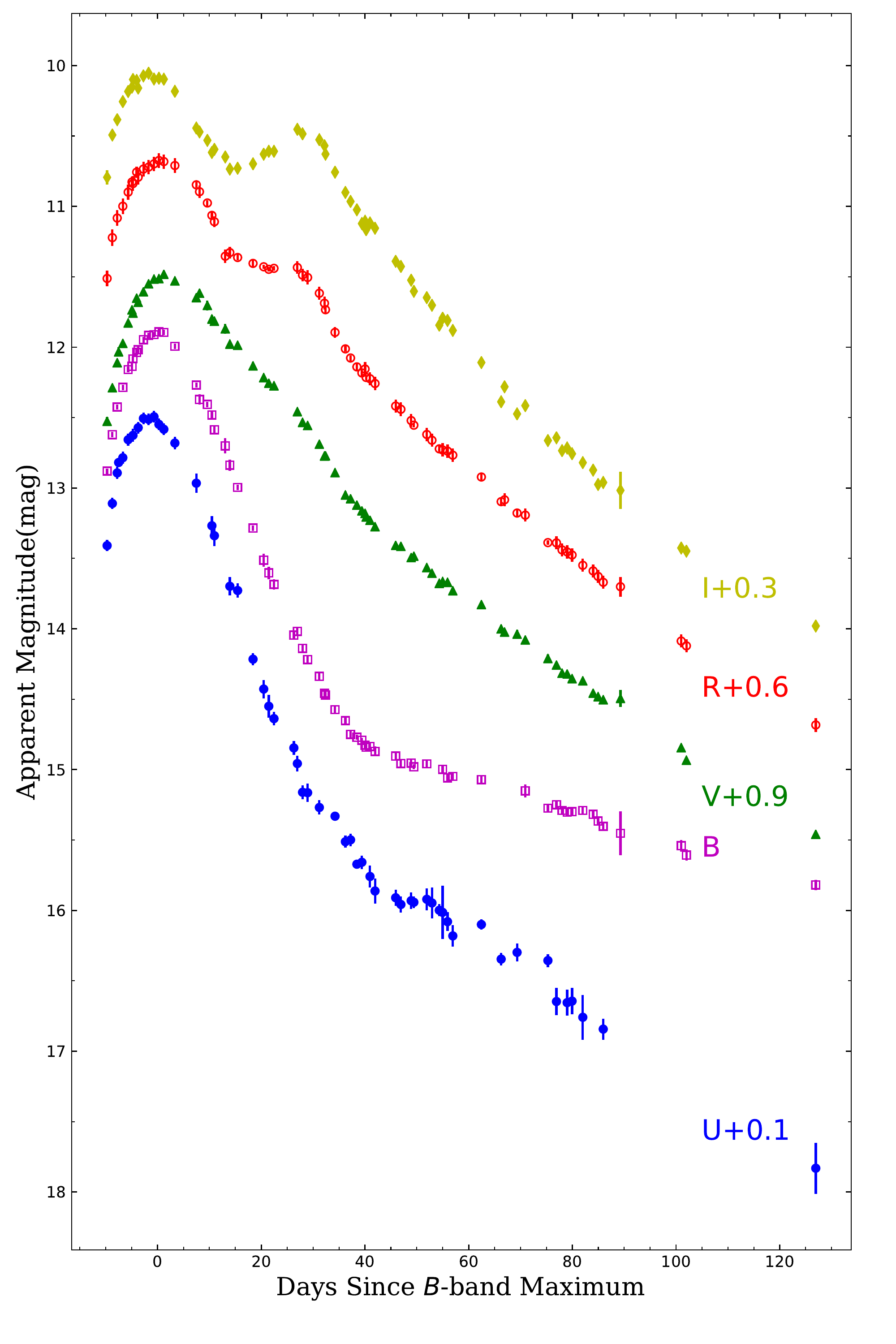}
\vspace{0.2cm}
\caption{The \textit{UBVRI} light curves of SN 2014J, spanning the first 5 months of evolution.}
\label{UBVRI} \vspace{-0.0cm}
\end{figure}

Figure \ref{UBVRI} shows our \textit{UBVRI} light curves of SN 2014J obtained during the first 5 months of evolution. From a low order polynomial fit to the near-maximum light curve, we derive $B_{\rm max}$ = 11.92$\pm$0.03 mag, $t_{Bmax}$ = JD 2456690.1$\pm$0.1 day, and $\Delta m_{\rm 15}(B)_{obs}$ = 0.98$\pm$0.01 mag. Our measurement is consistent with the result reported in \cite{2014MNRAS.443.2887F} and \cite{2016MNRAS.457.1000S}. However, SN 2014J suffers from large reddening which shifts the effective wavelength red-wards. The intrinsic decline rate parameter derived by  \cite{1999AJ....118.1766P} is 
\begin{center} $\Delta m_{\rm 15}(B)_{true}$ $ \simeq $ $\Delta m_{\rm 15}(B)_{obs}$ + 0.1 $\times$ $E(B-V)_{obs}$
\end{center}

According to \cite{2011ApJ...737..103S},  the Galactic reddening towards M82 is $E(B-V)_{MW}$ = 0.138 mag. However, as pointed by \cite{2009ApJS..183...67D}, this estimate may be strongly contaminated by point source emission from M82 itself. They suggested a lower value of $E(B-V)_{MW}$ = 0.059 mag on the \cite{1998ApJ...500..525S} scale, which is close to the value used by \cite{2015MNRAS.453.3300A}. Converting to the \cite{2011ApJ...737..103S} scale with a factor of 0.86, the final Galactic reddening adopted in our analysis is $E(B-V)_{MW}$ = 0.052 mag with
the classic reddening law of R$_V$ = 3.1 \citep{1989ApJ...345..245C}, consistent with that used by \cite{2014MNRAS.443.2887F}. 

Several studies discussed the total reddening towards SN 2014J but did not come to a consistent conclusion (see below). In our study, we adopt a model-independent reddening $E(B-V)_{obs}$ =1.19 $\pm$ 0.14 mag with a reddening law of $R_V = 1.64 \pm 0.16$, as in \cite{2014MNRAS.443.2887F}. Therefore, the reddening-corrected $\Delta m_{\rm 15}(B)_{true}$ is $\simeq $ 1.10 $\pm$ 0.02 mag for SN 2014J, similar to those of SN 2003du (1.04 $\pm$ 0.02 mag, \citealt{2005A&A...429..667A}; 1.02 $\pm$ 0.05 mag, \citealt{2007A&A...469..645S}),
SN 2005cf \citep[1.07 $\pm$ 0.03 mag,][]{2009ApJ...697..380W}, and SN 2011fe \citep[1.18 $\pm$ 0.03 mag,][]{2016ApJ...820...67Z}. 

\begin{figure}[htbp]
\center
\subfigure[\textit{B} band]{%
\includegraphics[width=.6\linewidth]{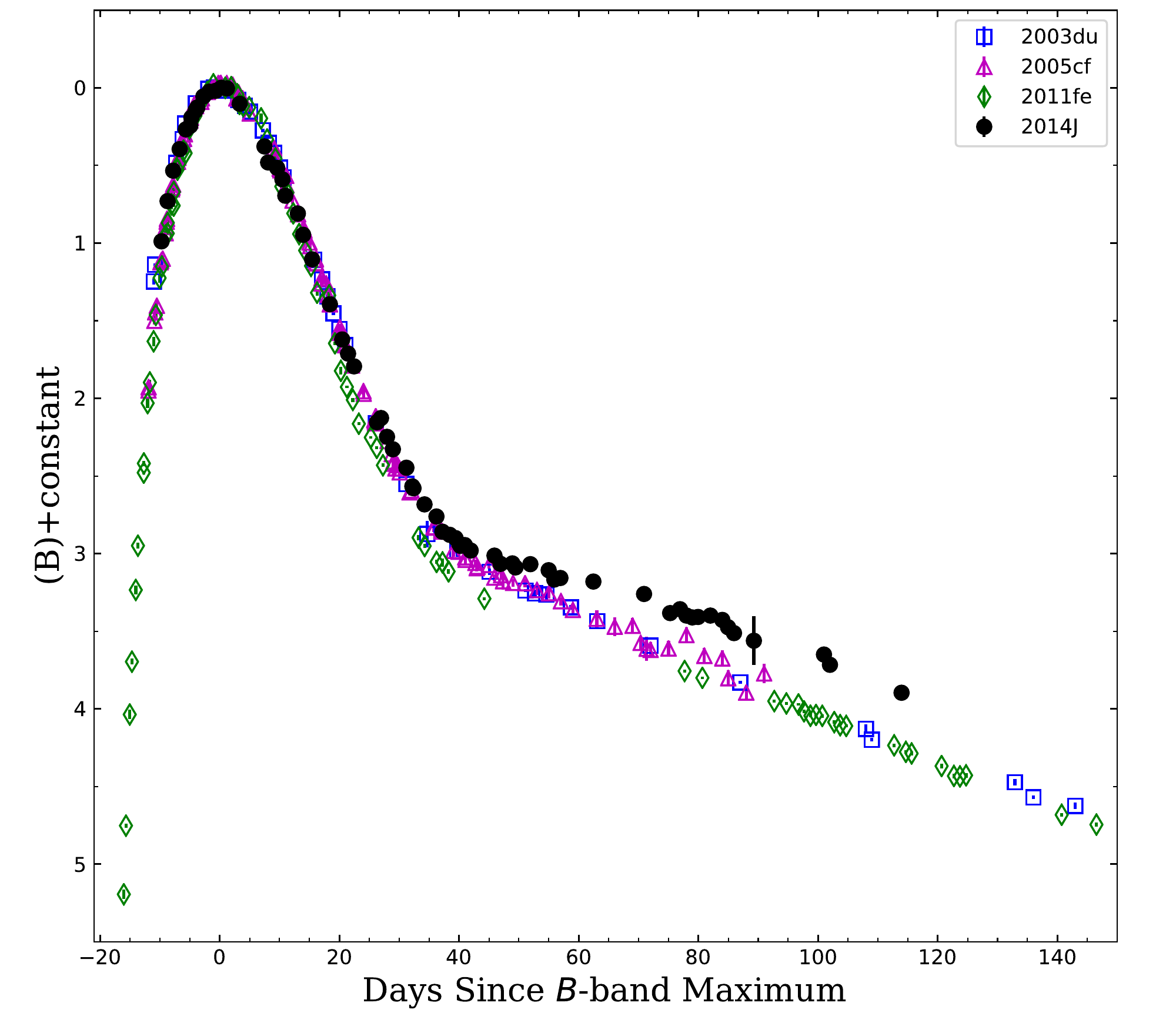}}
\subfigure[\textit{V} band]{%
\includegraphics[width=.6\linewidth]{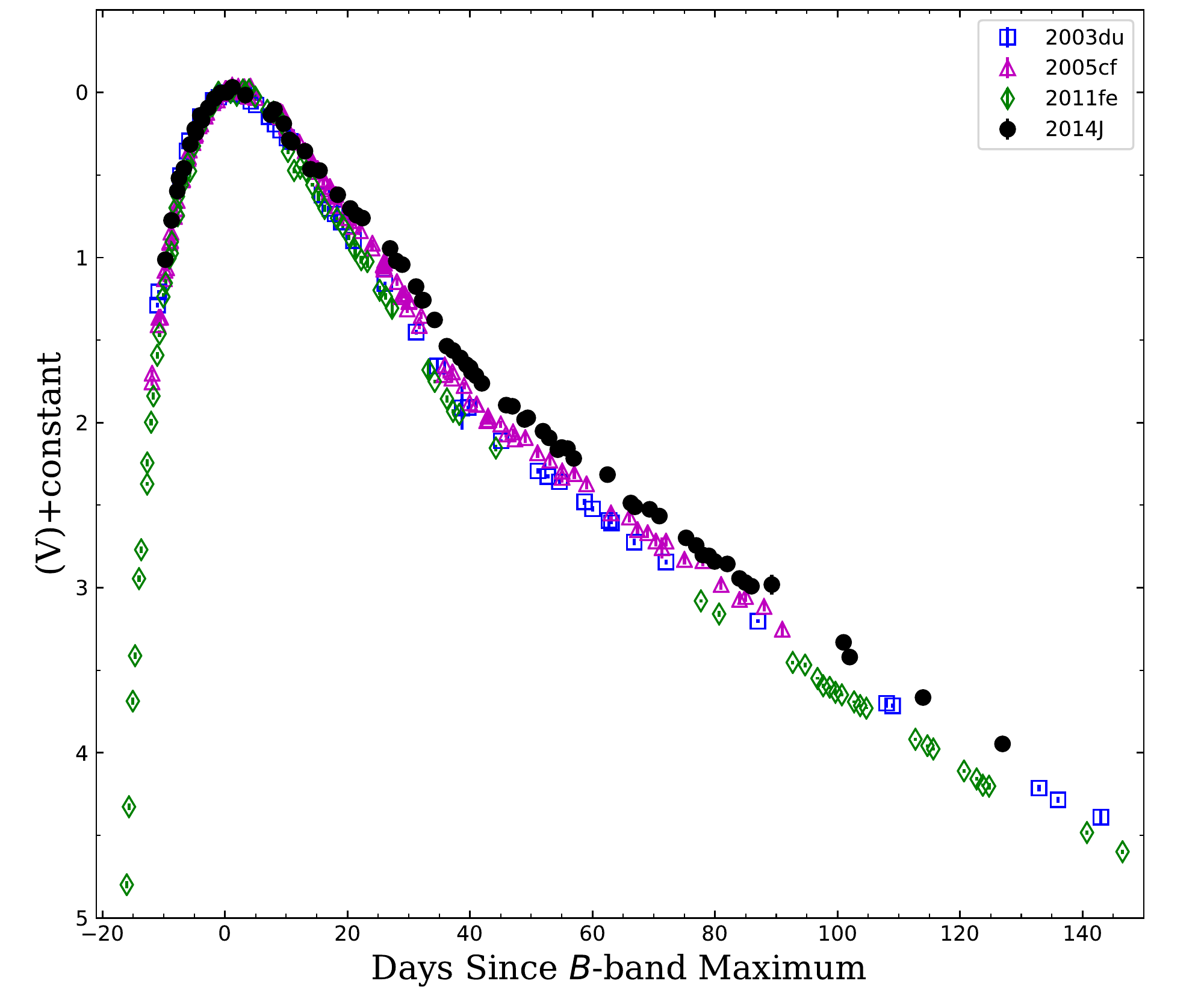}}
\vspace{0.2cm}
\caption{Comparison of the light curves of SN 2014J with SNe 2003du, 2005cf and 2011fe. The magnitudes are normalized to the peak and the phases are shifted to $t_{Bmax}$ for each SN.}
\label{B} \vspace{-0.0cm}
\end{figure}

Figure \ref{B} shows the comparison of the \textit{B}- and \textit{V}-band light curves of SN 2014J with those of SN 2003du, SN 2005cf, and SN 2011fe. 
We notice that the \textit{B}-band light curve of SN 2014J becomes somewhat flatten at t$\sim$+40 days, and shows emission excess relative to the compared SNe Ia with similar $\Delta$m$_{15}$(B). The deviation in the  \textit{B} band is evident, while this trend is less prominent in the \textit{V}-band. The decline rates measured during the period t$\sim$50-110 days are tabulated in Table \ref{decline}. Previous works suggest that SN 2014J is a normal SN Ia, except for the higher velocity of the ejecta \citep{2018MNRAS.481..878Z}, this flattening effect may be related to dust scattering of the SN light, as suggested by \citet{2018arXiv181011936W}. 
                                                          
In the following analysis, we attempt to model the $B$-band emission excess with additional SN photons scattered back by dusty medium, i.e. a light echo. In this scenario, we adopt a single scattering approximation following the procedure described in \cite{2005MNRAS.357.1161P}. If $d$ is the distance between SN 2014J and the observer, we can assume that $c{\Delta}t \ll d$, where $c$ is the speed of light and $\Delta$t is the duration of the SN radiation. As $d$ is much larger than other geometrical quantities considered here, at any moment, the distribution of the scattered photons along the  line of sight  can be approximated as a paraboloid with the SN locating at its focus. $L_{\lambda}(t)$ is the luminosity of the SN at wavelength ${\lambda}$ and $F_{SN,\lambda} = L_{\lambda}(t)/4{\pi}d^2$ is the corresponding flux. Thus the flux of the scattered light at a given delay time $t$ is:  
\begin{equation}
F_{\lambda}(t) = \int_0^tF_{SN,\lambda}(t-t') f(t')\,dt'\
\label{flux}
\end{equation}
In the above expression, the kernel function $f(t)$ includes all the physical and geometrical information of the dust and is expressed as:
\begin{equation}
f(t) = C_{ext}{\omega}c\int_{-ct/2}^{+\infty}\frac{{\Phi}(\theta)}{r}\int_0^{2{\pi}}n({\varphi}, z, t)\,d\varphi  dz\
\label{kernal}
\end{equation}  
where $\varphi$ and $z$ are defined as in Fig. 1 of \cite{2005MNRAS.357.1161P},  $C_{ext}$ is the extinction cross section, $\omega$ is the dust albedo, $\theta$ is the scattering angle, $n$ is the number density of the dust particles, $r (=z + ct)$ represents the distance between the SN and the dust volume element and ${\Phi}(\theta)$ is the scattering phase function satisfying $\int_{4\pi}{\Phi}(\theta)d\Omega = 1$. Both $\omega$ and ${\Phi}(\theta)$ are wavelength-dependent \citep{2001ApJ...548..296W,2003ApJ...598.1017D}.

To obtain the total flux received by the observer, the $F_{\lambda}(t)$ is added to the flux coming from the SN directly:
\begin{equation}
F_{T,\lambda}(t) = \left[ F_{SN,\lambda}(t)e^{-{\tau}_d} + F_{\lambda}(t)\right]e^{-{\tau}_I}
\label{total}
\end{equation}
Here ${\tau}_d$ is the optical depth of the CSM along the  line of sight , while ${\tau}_I$ is the optical depth of the ISM.

\subsection{Analytical Fitting}\label{gaussian kernel}
We apply the above model to the light curves of SN 2014J. The shape of $F_{SN,\lambda}(t)$ can be approximately represented by a flux distribution of a unreddened SN with similar $\Delta m_{\rm 15}(B)$ and spectral evolution. \cite{2016MNRAS.457.1000S} found that SN 2014J shares similarity with SN 2003du, except that the latter suffers negligible extinction \citep{2005A&A...429..667A}. Therefore we use the $B$-band light curve of SN 2003du, after shifting the peak magnitude to match the corresponding values of SN 2014J, as input light curve template to calculate $F_{B}(t)$. $F_{T,\lambda}(t)$ is the flux distribution of SN 2014J by definition. 
Assuming that the equation (\ref{kernal}) is Gaussian, the kernel function $f(t)$ can be simplified as: 
\begin{equation}
f(t) = \frac{A}{\sigma}exp\left[-\frac{(t-\tau)^2}{2\sigma^2}\right]
\label{gaussian}
\end{equation}
where $\tau$ represents the time delay of arrival of scattered light compared to direct light
, which is related to the distance and distribution of the dust. The standard deviation $\sigma$ is related to the physical size of the dust region responsible for the light echo. $A$ is the scale factor to measure the strength of the light echo. We substitute equation (\ref{gaussian}) into equations (\ref{flux}) and (\ref{total}) and then fit $t$ in the range of $[-8, 150]$ days\footnote{In our method, the exponential term of equation (\ref{total}) cannot be decoupled with the flux term (i.e., $F_{SN,\lambda}(t)$ and $F_{\lambda}(t)$).}. 

\begin{figure}[htbp]
\center
\includegraphics[width=\linewidth]{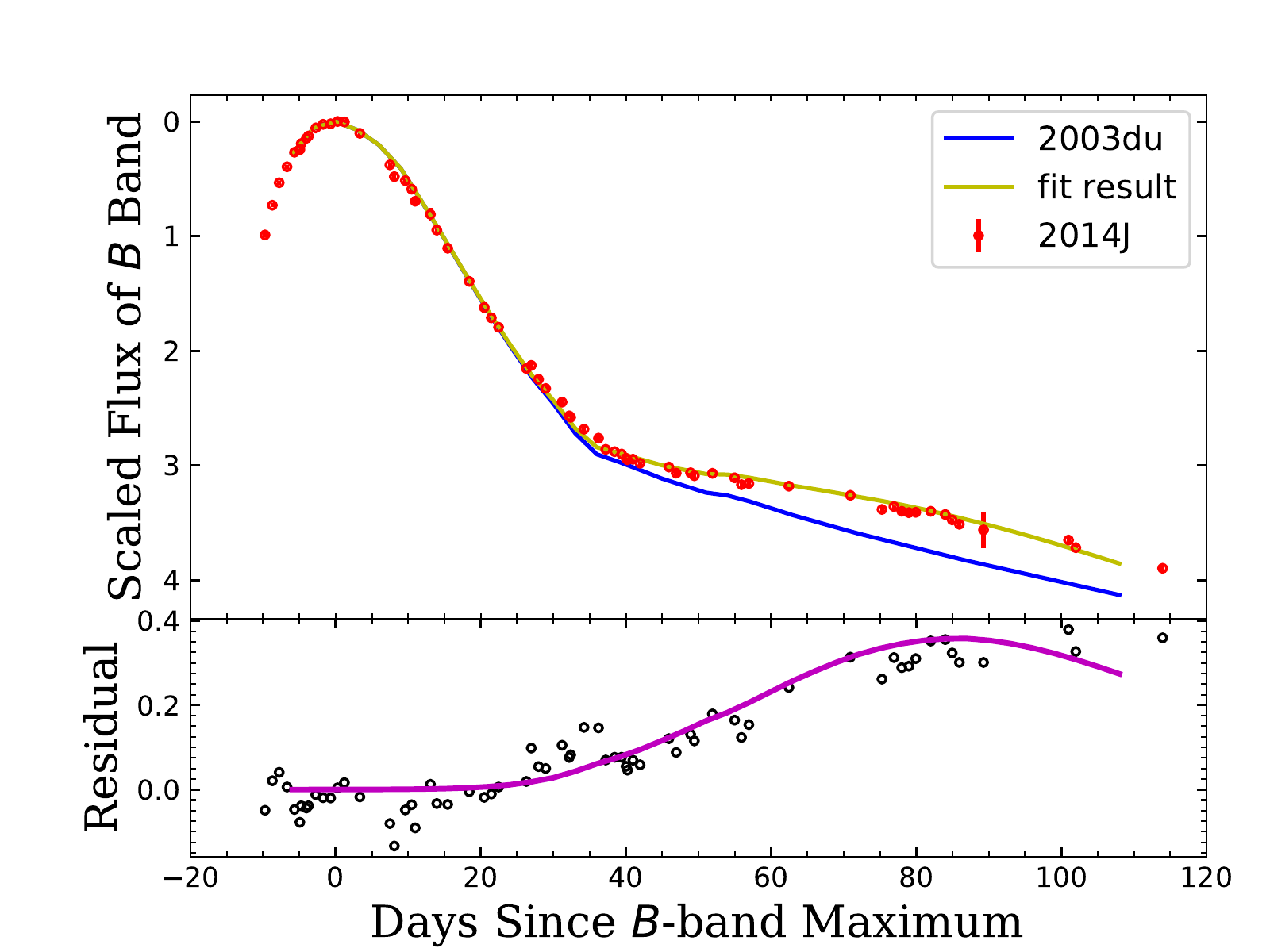}
\vspace{0.2cm}
\caption{The best fit to the normalized $B$-band light curve of SN 2014J based on the model of dust scattering. The light curves of SN 2014J and SN 2003du are shown (red dots and solid blue curve, respectively). The yellow line represents the light curve of SN 2003du  including the additional contribution of photons scattered by the dust shell located at $\sim 10^{17}$cm.}
\label{fit result1} \vspace{-0.0cm}
\end{figure}

The best-fit parameters and 1-$\sigma$ error listed in Table \ref{fit} are $A=0.12\pm0.03$, $\tau=64 \pm 8$ days and $\sigma=26\pm8$ days. Hence, we obtain $c\tau =$ 64 $\pm$ 8 light days (or $\approx$ 1.7 $\pm$ 0.2$\times$10$^{17}$ cm). Figure \ref{fit result1} shows that the light echo begins to emerge at around 20 days and increases to the peak value (${\sim} 30\%$ of total flux) at around 90 days. The distance between the SN and the center of the dust shell is calculated as:

\begin{equation}
R = \frac{c\tau}{1-cos\theta}
\label{distance}
\end{equation} 

Thus the dust distance from the SN can be estimated as $R>c\tau/2=$ 8.3 $\pm$ 1.0 $\times$10$^{16}$ cm.

More accurate estimation of distance relies on the information about the structure of the dust responsible for the scattering. In our analysis we consider two scenarios in our analysis, spherical shell and disk-like geometry (see Section \ref{monte}, for the latter case). We first assume that the dust cloud is a spherical shell around the SN. According to the fitting result, $f(t>100 days)\ll10^{-2}$ day$^{-1}$, the outer boundary of dust shell is thus inferred as $r_o$ $<$ 100 light days (or $<$ 2.6 $\times$10$^{17}$ cm). Therefore the dust should be of CS origin. \cite{2011ApJ...735...20A} suggest that the minimum radius for the CS dust around SNe Ia should be $\sim10^{16}$ cm because of radiant evaporation. We use the mass limit from \cite{2017MNRAS.466.3442J} to calculate the inner boundary radius $r_i$ of the dust shell. The dust shell should be formed by stellar winds and its density $\rho\propto\frac{\dot{M}_{wind}}{v_{wind}r^2}$. Given a mass loss rate $\dot{M}_{wind}$ $\sim10^{-7}$ M$_\odot$ yr$^{-1}$ \citep{2007ApJ...663.1269N} and a stellar wind velocity $v_{wind}$ $\sim$ 100 km s$^{-1}$, the mass of CSM can be expressed as 
\begin{center}
 $M_{CSM}=\int_{r_i}^{r_o}\rho dV=\int_{r_i}^{r_o}\frac{\dot{M}_{wind}}{v_{wind}r^2}4\pi r^2dr\approx 4 \times 10^{-21}(r_o-r_i)$ M$_\odot$ cm$^{-1}$
\end{center}
 where, $r_i$ is the radius of the inner boundary of the dust shell. The analysis of \cite{2017MNRAS.466.3442J} suggests that the pre-existing CSM has a mass $M_{CSM}\le 10^{-5}$ M$_\odot$ within $r\sim 10^{17}$cm. That requires $(r_o-r_i)\le 2.5\times 10^{15}$cm, consequently, implying that the dust shell should be very thin. However, putting these estimates and the typical values of $C_{ext}$, $\omega$, ${\Phi}(\theta)$ \citep{2001ApJ...548..296W,2003ApJ...598.1017D} and $n$ into equation (\ref{kernal}), the resultant $f(t)$ would be too small to produce a significant light echo as seen in SN 2014J. Thus, a spherical dust shell seems unlikely for SN 2014J.

\subsection{Monte Carlo Fit}\label{monte}

We run a Monte Carlo simulation to fit the $B$-band light curves of SN 2014J. Mie scattering theory \citep{1908AnP...330..377M} is used in the calculations of the scattering process (Hu in prep.). Adopting the distribution derived by \cite{2015ApJ...811L..39N} (assuming an average radius for the dust grain of about 0.036 $\mu m$), the size distribution of the dust is expressed as 
\begin{equation}
f(r) = r^{-a_0}exp{[-b_0(log\frac{r}{r_0})^{2.0}]}
\end{equation}
 where $r$ is the radius of a dust grain, ranging from 5 to 500 nm. We adopt ${a_0} = 4.0, b_0 = 7.5 $ and $r_0 = 0.04~\mu m$ in order to give size distribution similar to \cite{2015ApJ...811L..39N}.

As for the geometric distribution, we consider a disk structure for the dust shell. There are five free parameters to describe the disk  \citep{2017ApJ...847..111N}, including the observing angle $\theta_{obs}$, the opening angle of the disk $\theta_{disk}$, the inner radius $R_{in}$, the width of the CSM $R_{wid}$, and the optical depth in the $B$ band $\tau_B$. The radial distribution of the dust density is assumed to have an index $n=$  $-$2 (i.e., $\rho _{dust}(r) = \rho _{dust}(r_{in})(r/r_{in})^{-2}$). The ranges of these parameters are in Table \ref{parameter}. We adopt silicate grains for our dust model. Following the procedure described in Section \ref{gaussian kernel}, the $B$- and $V$-band light curves of SN 2003du were used as templates in the calculations. 

There is a degeneracy between different combinations of the above parameters, i.e., different sets of parameters produce similar light curves. Thus, we list in Table \ref{fit result2} four sets (two groups, Disk 1 and Disk 2 with $\theta_{disk}$=15$^{\circ}$ and 30$^{\circ}$, respectively) of parameters, producing light curves similar to SN 2014J. 

\begin{figure}[htbp]
\center
\subfigure[\textit{B} band]{%
\includegraphics[width=.45\linewidth]{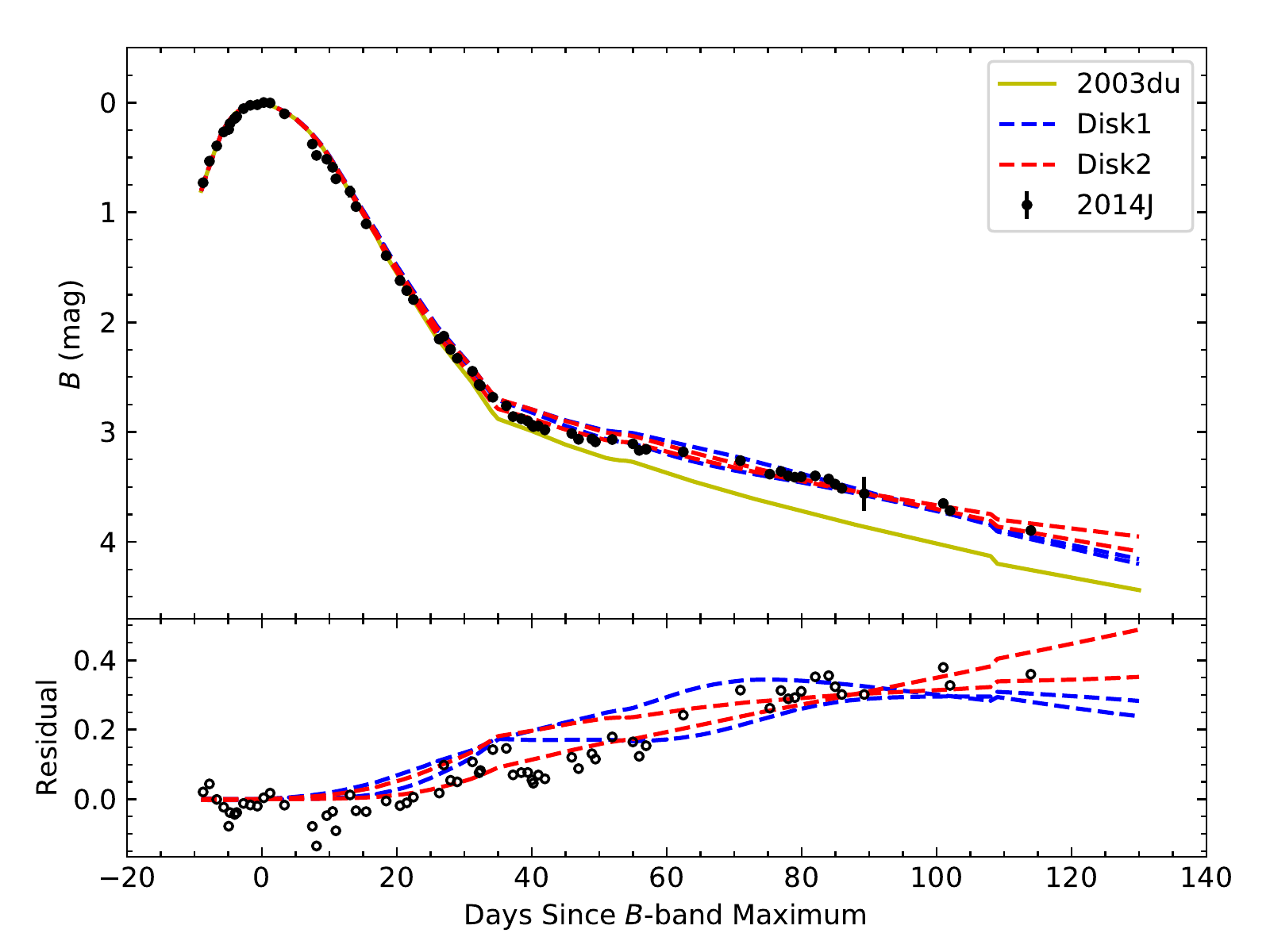}}
\subfigure[\textit{V} band]{%
\includegraphics[width=.45\linewidth]{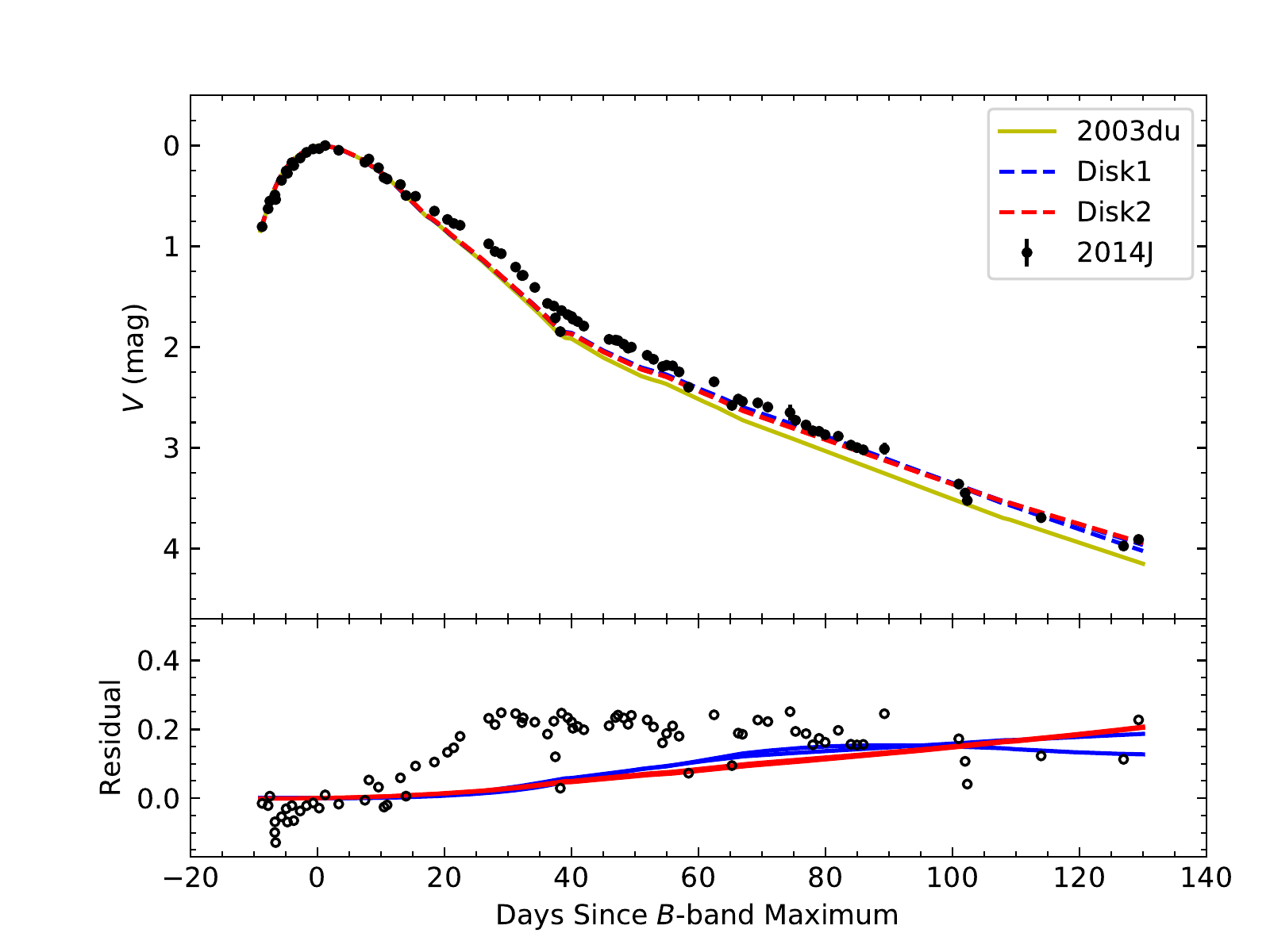}}
\subfigure[\textit{R} band]{%
\includegraphics[width=.45\linewidth]{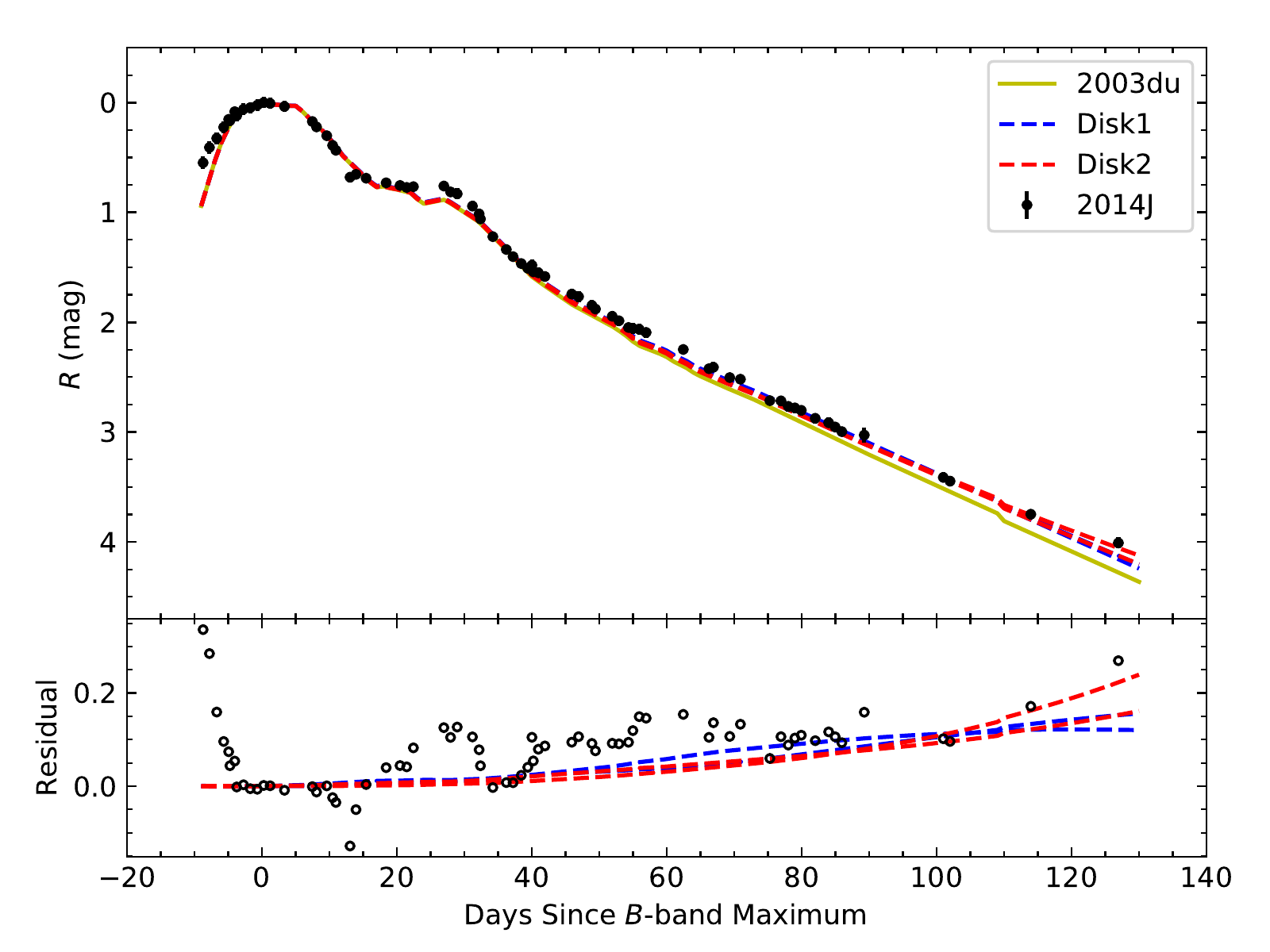}}
\subfigure[\textit{I} band]{%
\includegraphics[width=.45\linewidth]{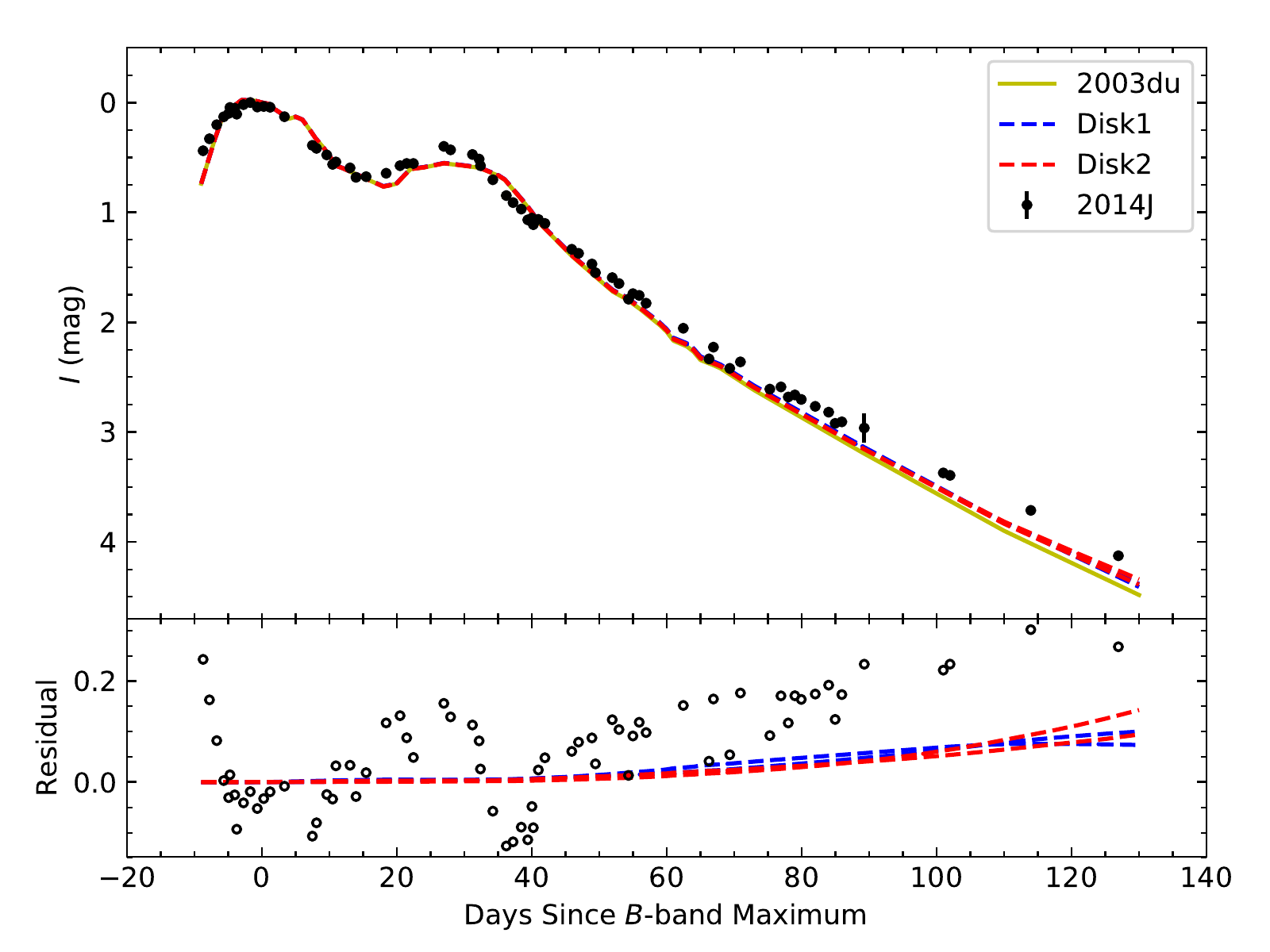}}
\vspace{0.2cm}
\caption{Comparison of the $BVRI$-band light curves of SN 2014J with those derived from Monte Carlo simulations. The black circles in the residual plot in each panel are the magnitude difference between SN 2014J and SN 2003du at similar phase, while the dash lines represent corresponding  magnitude difference between the simulated curves (after considering the effect of dust scattering) and the observed values of SN 2003du.}
\label{MC} \vspace{-0.0cm}
\end{figure}

The $B$-band observed light curve of SN 2014J and the best-fit model are shown in Figure \ref{MC}. The flux excess in the $B$-band light curve is well fitted by our simulation. The scattered SN light emerges after t$\sim$+20 days, and becomes progressively stronger (Disk 2), reaching the peak around +80 $sim$ +100 days (Disk 1).
However, in the $V$-band, the scattering effect is much smaller than that seen in the $B$ band. The magnitude difference between +10 days and +80 days of these two SNe Ia cannot be explained by our model. The scattered light barely affects the light curves at longer wavelengths, as shown in Figures \ref{MC}(c) and (d). Therefore, we suggest that the diversity of the two SNe in $R$ and $I$ bands may be mainly due to their intrinsic specificity. However, we caution that the high degeneracy of our model prevents us from analyzing more quantitatively the CSM distribution. The CSM for such a geometry can be formed when the wind of the red giant companion concentrates on the  equatorial plane.
However, \cite{2014ApJ...790...52M}, \cite{2014ApJ...790....3K}, \cite{2014ApJ...792...38P} and \cite{2014ApJ...784L..12G} found the companion of SN 2014J is unlikely to be a red giant star with steady mass transfer or a luminous symbiotic system such as RS Ophiuchi \citep{2012Sci...337..942D}. Fainter recurrent novae are still possible candidates for the progenitor system of SN 2014J.
On the other hand, some variations of the DD scenario might also form disk-like dusty CSM, e.g. mass outflows during rapid accretion during the final evolution \citep{2010ApJ...709L..64G,2011ApJ...737...89D} and magnetically driven winds from the disk around the WD-WD system \citep{2013ApJ...773..136J}.

\section{Nebular Phase Evolution}\label{late}
\begin{figure}[htbp]
\center
\subfigure[\textit{HST} WFC3/UVIS F438W images]{%
\begin{minipage}[b]{1.\linewidth} 
\centerline{\includegraphics[width=.8\linewidth]{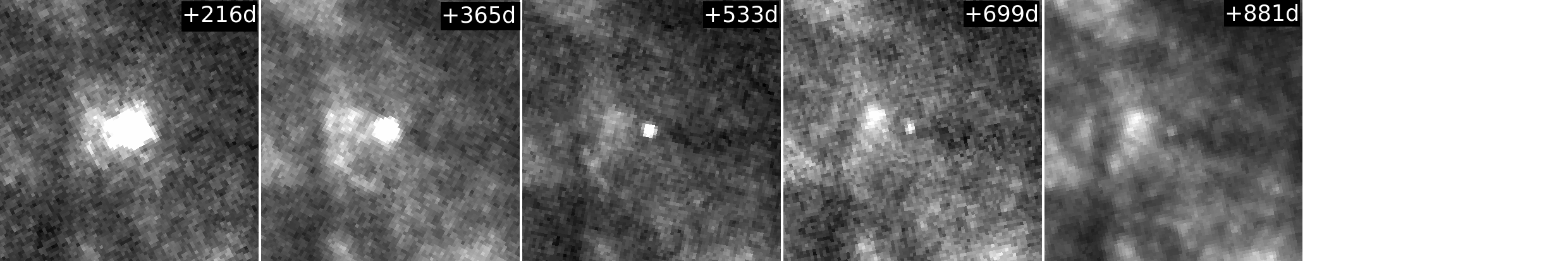}}
\end{minipage}}
\subfigure[\textit{HST} WFC3/UVIS F555W images]{%
\begin{minipage}[b]{1.\linewidth} 
\centerline{\includegraphics[width=.8\linewidth]{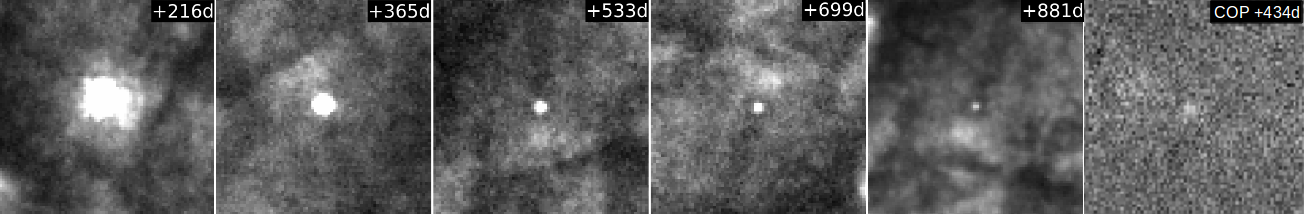}}
\end{minipage}}
\caption{\textit{HST} images of SN 2014J. We label the corresponding phase at the top right corner of each image. The SN can be clearly seen in the images taken as late as +881 days after the $B$-band maximum. The light echo is visible in the bottom left quadrant of the F438W-band images taken on +365 days and +533 days. We also display the central part of the latest ground-based, template-subtracted $V$-band image taken with the COP 1.82~m telescope on ~+434 days in the F555W sequence.}
\label{f555} \vspace{-0.0cm}
\end{figure}

\begin{figure}[htbp]
\centering
\includegraphics[width=\linewidth]{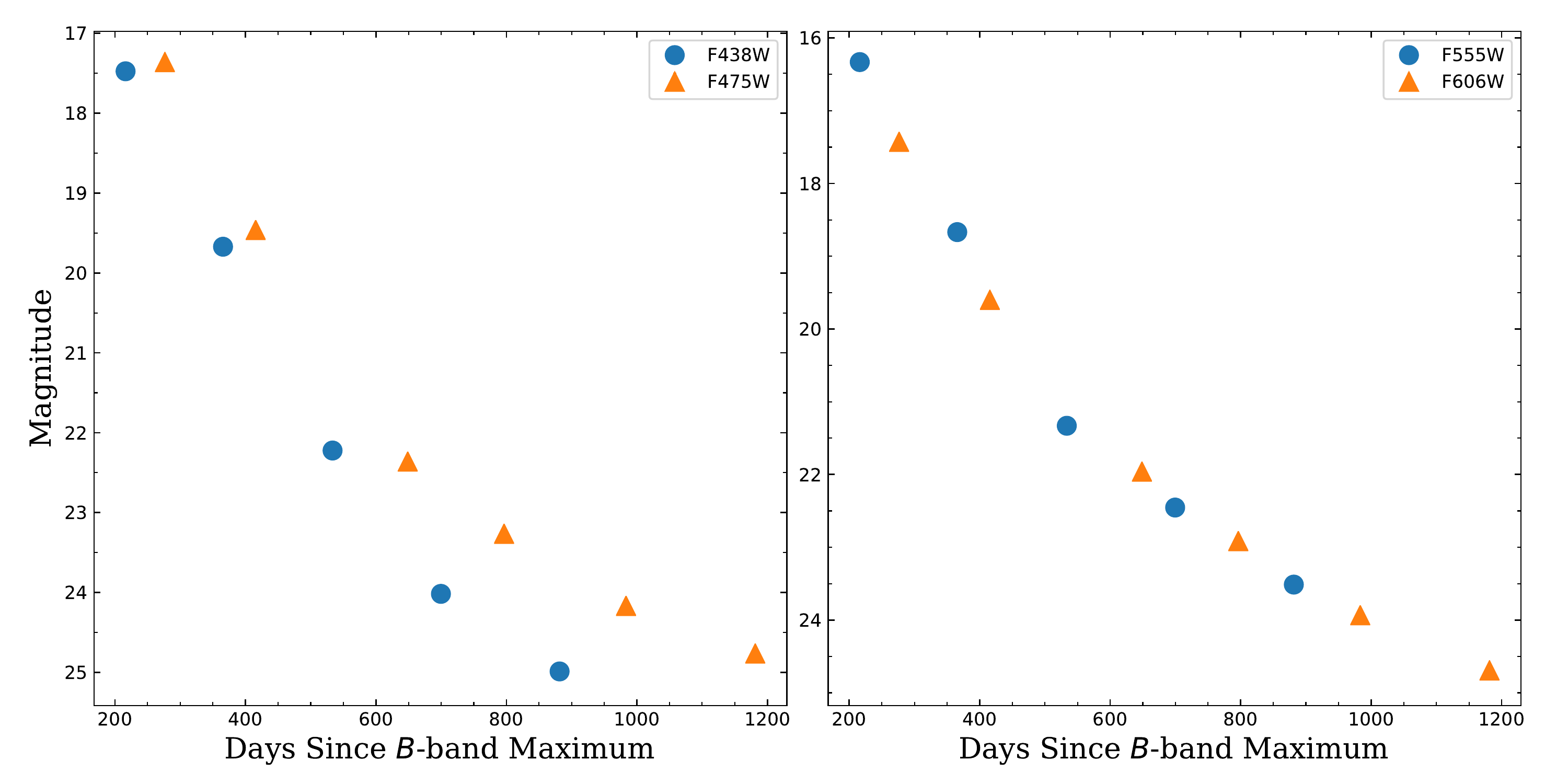}

\caption{Late-time \textit{HST} photometry of SN 2014J. Left: comparison of the WFC3/UVIS $F438W$- and ACS/WFC $F475W$-band photometry. Right: comparison of the WFC3/UVIS $F555W$-band and ACS/WFC $F606W$-band photometry.}
\label{HST_photo} 
\end{figure}

We  examined the very late time evolution of SN 2014J, based on the $HST$ observations. The images taken with the $HST$ WFC3/UVIS in F438W and F555W bands, centering at the position of SN 2014J, are chronologically displayed in Figure \ref{f555}. In Figure \ref{HST_photo}, we compare F438W- and F555W-band photometry with the F475W- and F606W-band photometry obtained with $HST$ ACS/WFC \citep{2018ApJ...852...89Y}. Our first two epochs photometry are consistent with the values reported in \cite{2017ApJ...834...60Y} \footnote{The remained discrepancies are mainly due to the difference between Vega and AB magnitude system.}. The $HST$ magnitudes obtained in the F555W and F606W bands are in agreement with each other from t$\sim$200 days to t$\sim$1000 days, while some discrepancy exists between the F438W- and F475W-band magnitudes. 

\begin{figure}[htbp]
\center
\includegraphics[width=\linewidth]{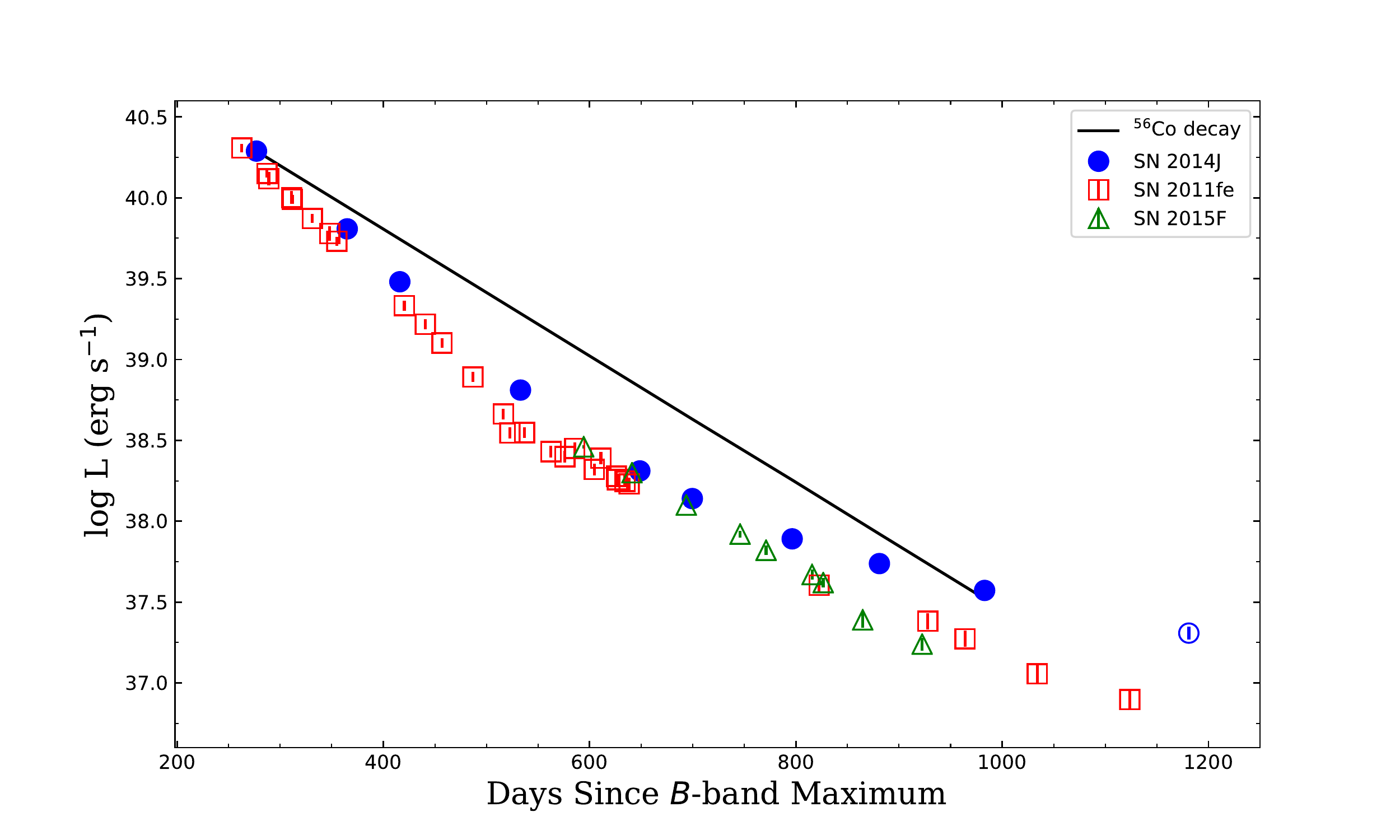}
\vspace{0.2cm}
\caption{Comparison of the luminosity evolution of SNe 2011fe, 2014J and 2015F. The black solid line displays the decline rate of pure $^{56}$Co decay. The last empty point of SN 2014J is taken from \cite{2017ApJ...834...60Y}.}
\label{luminosity} \vspace{-0.0cm}
\end{figure}

We combined our data taken after t$\sim$ 300 days with those obtained with the $HST$/WFC mentioned above to construct the pseudo-bolometric light curve in the wavelength range from $\sim$3500{\AA} to $\sim$9000{\AA}, following the same procedure as \cite{2018ApJ...852...89Y} with the warped spectra (see step (5a) in \cite{2018ApJ...852...89Y} Section 3.1). The pseudo-bolometric light curve is shown in Figure \ref{luminosity}. The spectral evolution of SN 2014J is similar to SN 2011fe, and there is only minor evolution between the spectra of SN 2011fe taken at +576 days \citep{2015MNRAS.454.1948G}  and +1016 days \citep{2015MNRAS.448L..48T} after $B$ band maximum. Therefore we use the spectrum of SN 2011fe at +1016 days to calculate the luminosity of SN 2014J for phases later than $\sim+500$ days after maximum. The bolometric light curve has been corrected for both the Galactic and host-galaxy extinction, and a distance of 3.53 Mpc \citep{2009ApJS..183...67D} is adopted in the calculation. The Galactic and host-galaxy extinction has been corrected (see Section \ref{early}). 
The decline rates measured at different phases for SN 2014J are listed in our Table \ref{decline rate}, consistent with Table 3 in \cite{2018ApJ...852...89Y}.
The pseudo-bolometric light curve of SN 2014J declines over 1.3 mag/100d, which is faster than the $^{56}$Co decay (i.e., 0.98 mag/100d), from $\sim$+300 days to $\sim$+500 days. After $\sim$+500 days, the decline rate decreases to reach 0.40 $\pm$ 0.05 mag/100d when approaching $\sim$+1000 days after the maximum. The fact that the very late-time light curve drops slower than the $^{56}$Co decay implies that there are other power sources besides radioactive decay energy from $^{56}$Co. In the following we analyze these two periods separately.

\subsection{Pseudo-Bolometric Light Curve between $\sim$+300 days and $\sim$+500 days}

As shown in Figure \ref{luminosity}, the bolometric light curve of SN 2014J declines faster than the $^{56}$Co decay. The late-time luminosity evolution of SN 2011fe \citep{2017MNRAS.468.3798D} and SN 2015F \citep{2018ApJ...859...79G} are also shown  for comparison. These two SNe also show faster decline rate compared to $^{56}$Co decay during this period. The last measurement (at 1181 day) in \cite{2017ApJ...834...60Y} is also plotted  in Figure \ref{luminosity}.  The luminosity evolution of SN 2014J shows a transition at t$\sim$+500 days, also reported for SN 2011fe \citep{2017MNRAS.468.3798D}. Before and after this period the light curve declines linearly but with a different rate, as shown in Figure \ref{luminosity}. 
Between $\sim$+200 days and $\sim$+500 days, SNe Ia are believed to be mainly powered by electron/positron annihilation \citep{2015MNRAS.454.3816C}. If the positrons are completely trapped by a magnetic field, the decline rate should follow the decay rate of $^{56}$Co. The faster decline rate can be explained by a weak or radially combed magnetic field \citep{1999ApJS..124..503M}. However, \cite{2017NatAs...1E.135C} argued that the positron annihilation signal observed in Milky Way requires a stellar source of positrons to have an age of 3-6 Gyr, strongly against  significant positron escape in normal SNe Ia. Additionally, previous analysis of late-time observations of SNe Ia does not favour positron escape \citep{2009A&A...505..265L,2014ApJ...796L..26K}. 

\begin{figure}[htbp]
\center
\includegraphics[width=\linewidth]{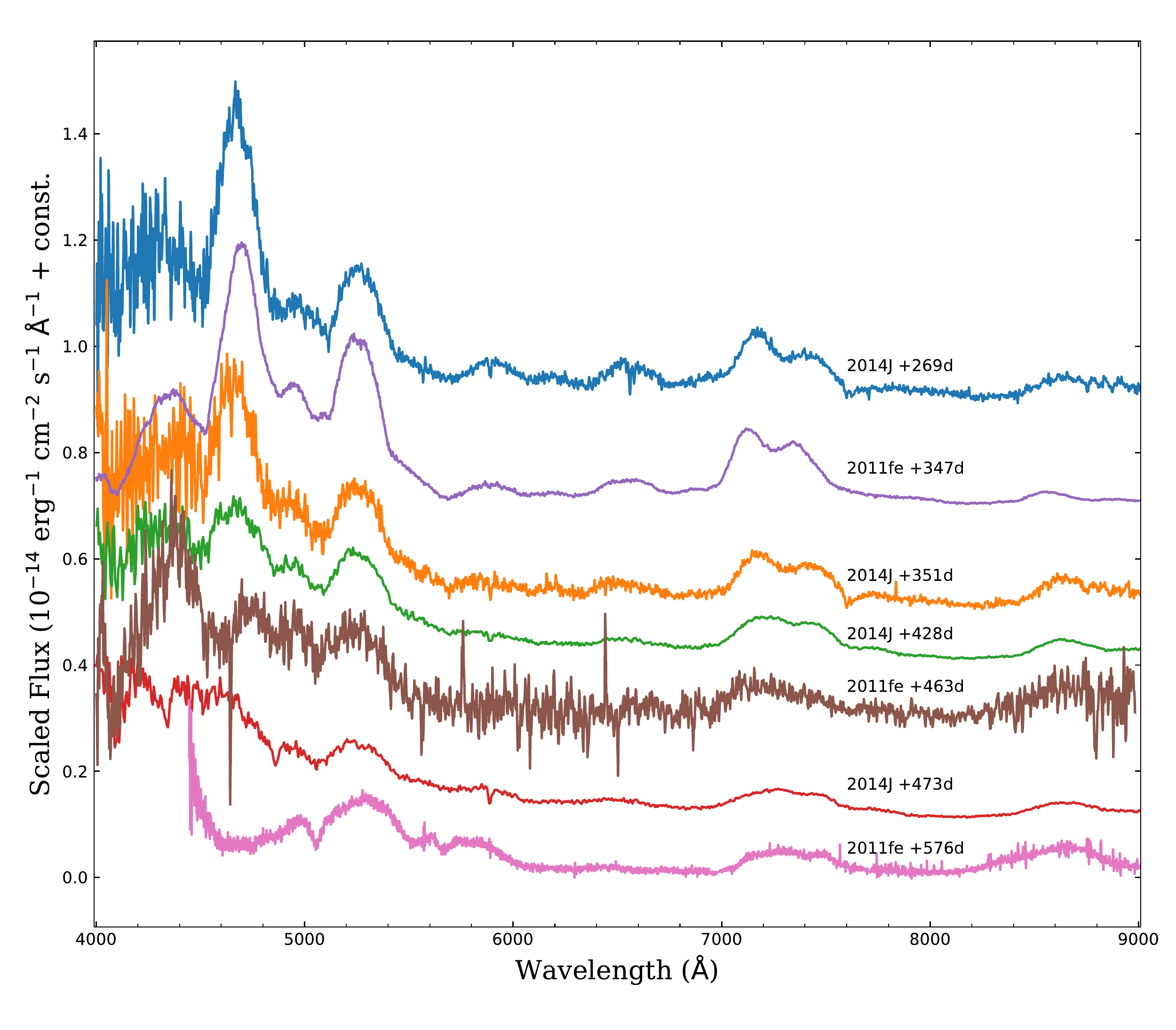}
\vspace{0.2cm}
\caption{Four late spectra of SN 2014J and three nebular spectra of SN 2011fe. For clarity, the spectra are arbitrarily shifted along the flux axis. The spectra data of SN 2014J  are in Data behind the Figure files.}
\label{late_spectra} \vspace{-0.0cm}
\end{figure}

An alternative explanation for this faster decline rate is the evolution of emission lines. The nebular-phase spectra of SN 2014J are available at t$\sim$ +269, +351, +428 and +473 days \citep{2016MNRAS.457.1000S, 2018MNRAS.481..878Z}, as shown in Figure \ref{late_spectra}. Overplotted are three nebular spectra of SN 2011fe, taken on +347 days \citep{2015MNRAS.450.2631M}, +463 days \citep{2016ApJ...820...67Z} and +576 days \citep{2015MNRAS.454.1948G}, respectively. All the spectra were downloaded from the WISeREP archive\footnote{http://wiserep.weizmann.ac.il/} \citep{2012PASP..124..668Y}. All spectra are flux-calibrated using the late-time ground-based and $HST$ photometry, with accuracy of about 0.03 mag. At these late phases, the main nebular emission features are blends around [Fe II] $\lambda$4400, [Fe III] $\lambda$4700 and [Fe II] $\lambda$5200. The $\lambda$4700 feature tends to become weak in both SN 2014J and SN 2011fe during the period from t$\sim$300 to t$\sim$500 days. In the +576d spectrum of SN 2011fe, this emission completely disappears. Considering the similar spectral evolution of SN 2011fe and SN 2014J, the $\lambda$4700 and the $\lambda$5200 features should vanish in SN 2014J at a similar phase. 

\begin{figure}[htbp]
\center
\includegraphics[width=\linewidth]{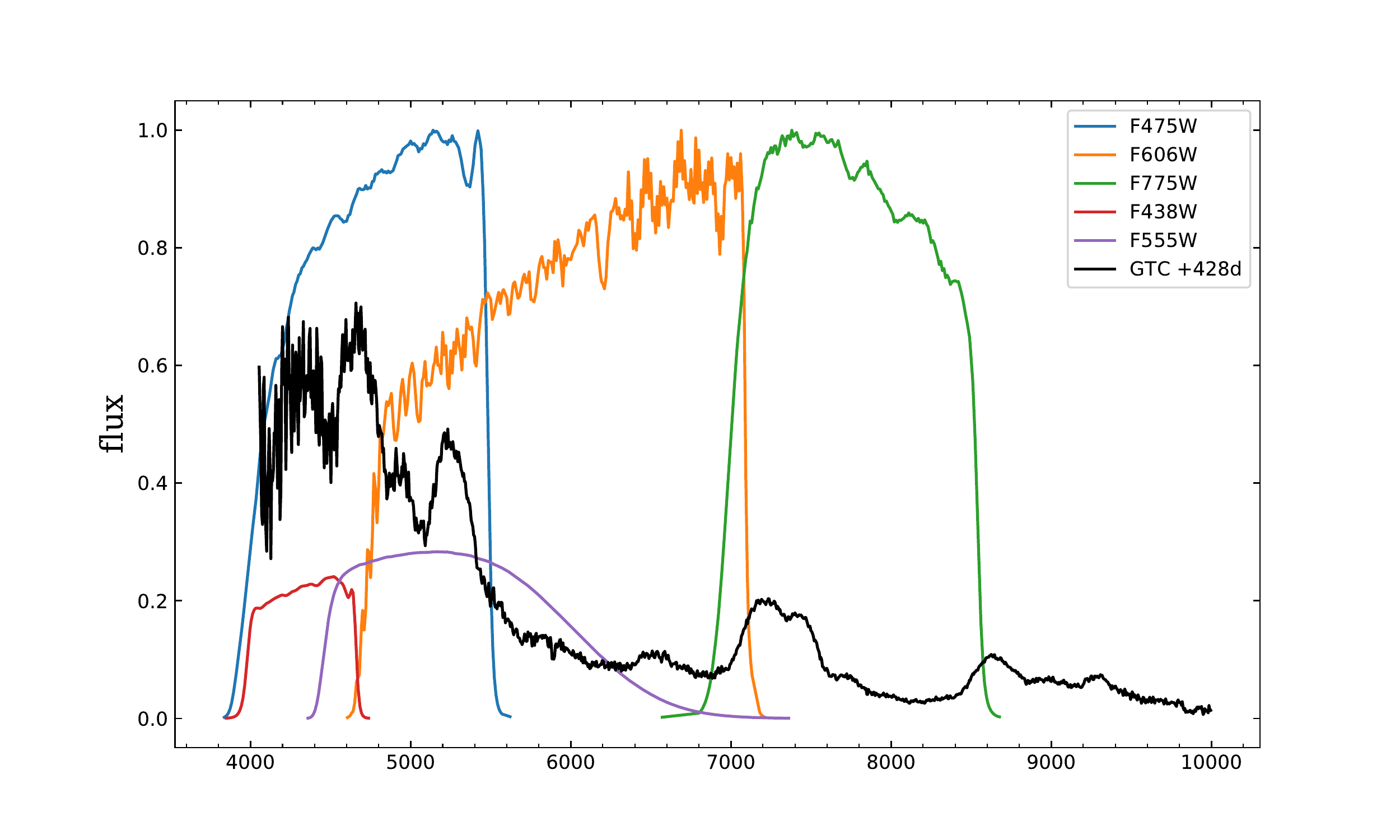}
\vspace{0.2cm}
\caption{The transmission curves of five $HST$ filters, shown with the +428d GTC spectrum of SN 2014J.}
\label{filter} \vspace{-0.0cm}
\end{figure}

In Figure \ref{filter} we over plot the transmission curves of five $HST$ filters with the +428 days spectrum, it is evident that the  ACS/WFC $F775W$ filter does not cover the fast-evolving spectral features. Thus the F775W-band light curve is expected to declines at a much slower rate than other bands during this period (see Table \ref{hphoto}). Therefore, the evolution of the prominent spectral features can affect the decline rate of the pseudo-bolometric light curves, which explains the faster decline of the bolometric light curves seen in SN 2014J and SN 2011fe. 
The ratio of $\lambda$4700 and $\lambda$5200 emission blends can be considered a proxy of ionization degree. The decrease in the ejecta temperature leads to a smaller ratio of $\lambda$4700/$\lambda$5200 
(and hence a lower ionization).

\subsection{Pseudo-Bolometric Light Curve after $\sim$+500d}
After t$\sim$500 days, the luminosity of SN 2014J declines more slowly than SN 2011fe and SN 2015F, as well as the $^{56}$Co decay, as also noticed by \cite{2018ApJ...852...89Y}. However, the complete trapping of positrons cannot explain this flattening in the light curve of SN 2014J. In addition to the decay of $^{56}$Co, several energetic mechanisms can affect the late-time evolution. For instance, 
leptonic energy from Auger and internal conversion electrons produced by the decay chain $^{57}$Co $\rightarrow$ $^{57}$Fe can slow down the late-time bolometric light curve of SNe Ia \citep{2009MNRAS.400..531S}. Besides, the ``freeze-out" effect can also have a similar consequence \citep{1993ApJ...408L..25F}. If there is a WD companion left after the supernova explosion, the energy released by the delayed radioactive decays of $^{56}$Ni and $^{56}$Co from  surface of the surviving WD can also contribute to the late-time SN light curve \citep{2017ApJ...834..180S}.
Nevertheless, it is not easy to distinguish different possible scenarios due to the lack of spectra after $\sim$ 500 days. Below we explore the possibility that the late-time flattening stems from long-term decay chain $^{57}$Co $\xrightarrow[\text{}]{\text{t$_{1/2}$=271.8d}}$ $^{57}$Fe and $^{55}$Fe $\xrightarrow[\text{}]{\text{t$_{1/2}$=999.7d}}$ $^{55}$Mn.

Since we have no infrared data at late time, we assume the pseudo-bolometric luminosity to be proportional to the bolometric luminosity. As the ejecta of supernova expand, the $\gamma$-ray optical depth decreases rapidly as $t^{-2}$, and can be neglected after 600 days \citep{2001ApJ...559.1019M}. Therefore our fit is restricted to data taken  after 650 days. Other power sources (light echo, light from the companion, etc.) are not included. The time-dependent contribution expression of each isotope with atomic number $A$ \{55, 56, 57\}  is \citep{2014ApJ...792...10S}:
\begin{equation}
L_{A}(t) = 2.221 \frac{{\lambda}_A}{A}\frac{{M}(A)}{\rm M\odot}\frac{q^l_Af_{Al}(t)+q^X_Af_{AX}(t)}{\rm keV}{\rm exp}(-\lambda_At)\times10^{43}\rm erg
\label{la}
\end{equation}
where ${{\lambda}_A=\frac{\rm ln(2)}{t_{1/2,A}}}$, $M(A)$ is the total mass of the element with atomic number $A$, $q^l_A$ and $q^X_A$ are the average energies carried by charged leptons and X-rays per decay, respectively. We adopt the values from Table 1 in \cite{2009MNRAS.400..531S}, including energies from Auger $e^-$, internal conversion $e^-$ and $e^+$. We assume full trapping of the charged leptons and X-rays, thus $f_{Al}=f_{AX}=1$. In Equation (\ref{la}) only the  \textit{M(A)} is a free parameter for each isotope. 

We then fit the light curve to get the $^{57}$Ni/$^{56}$Ni mass ratio and compare it with simulations in the literature. Firstly, we regard the mass of all three isotopes as free parameters. The derived $^{57}$Ni/$^{56}$Ni ratio is $0.124 \pm 0.041 $, which is 5 times the solar value,  ($^{57}$Fe/$^{56}$Fe)$_\odot=0.023$ \citep{2009ARA&A..47..481A}. This value is also much larger than that given by 2D \citep[0.032-0.044;][]{2010ApJ...712..624M}, 3D delayed detonation models \citep[0.027-0.037;][]{2013MNRAS.429.1156S} and the violent merger models \citep[0.024;][]{2012ApJ...747L..10P}. Additionally, the derived mass of $^{56}$Ni, $\sim$ 0.18 M$\odot$ is significantly lower than the typical value of SNe Ia (i.e., about 0.6 M$\odot$). This means that the full trapping of charged leptons is not a realistic hypothesis.

\begin{figure}[htbp]
\center
\includegraphics[width=\linewidth]{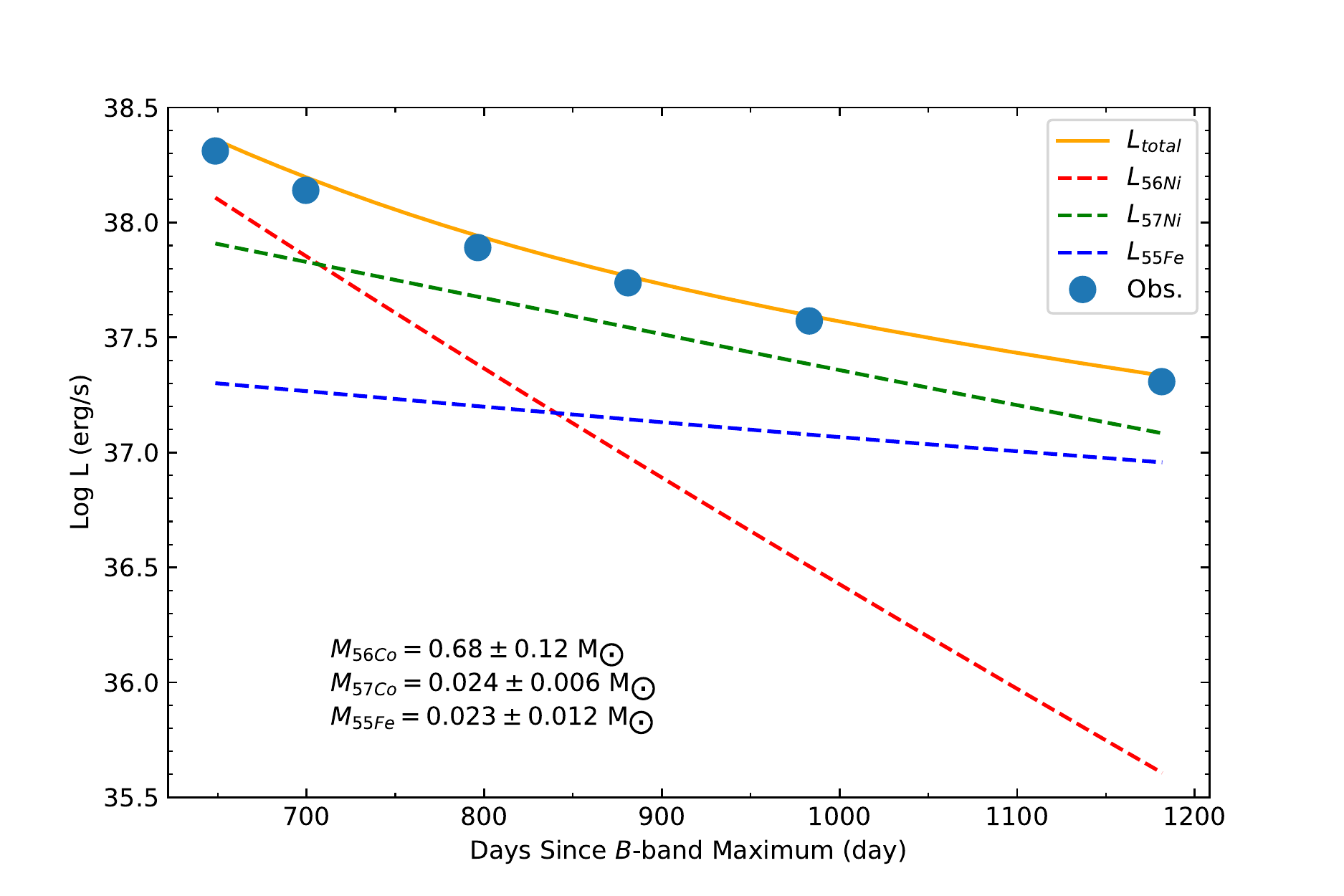}
\vspace{0.2cm}
\caption{Radioactive decay fits to the late-time pseudo-bolometric light curve of SN 2014J (open blue circles). The dash lines are the light curve contribution of three isotopes. The solid line is the sum of three isotopes luminosity.}
\label{555657} \vspace{-0.0cm}
\end{figure}

We account for lepton escape  by adopting $f_{Al}=1-{\rm exp}[-(\frac{t^l_A}{t})^2]$, where $t^l_A$ is the time when the optical depth for leptons  becomes unity.  We choose $t^l_{56}=249$ days and $t^l_{57}=812$ days from Case 1 in \citet{2017MNRAS.468.3798D}. We take $t^l_{55}=t^l_{57}$ because $^{55}$Fe and $^{57}$Ni both decay without production of positrons  \cite[see Table 1 in][]{2009MNRAS.400..531S}. The final fitting result is displayed in Figure \ref{555657}. Our updated fitting gives M$_{^{56}Ni}$ $=0.68 \pm 0.12$ M$\odot$
and $^{57}$Ni/$^{56}$Ni $=0.035 \pm 0.011$. Our estimate of the mass of $^{56}$Ni is fully consistent with the results from \citet{2016MNRAS.457.1000S}, \cite{2015ApJ...798...93T} and \cite{2014Natur.512..406C}. While the $^{57}$Ni/$^{56}$Ni ratio is about half of the value reported in \cite{2018ApJ...852...89Y}, although they assume no leptons escape and fixed $^{57}$Ni/$^{55}$Ni $=0.8$ in their calculations. Our new result is consistent with both 2D and 3D delayed-detonation simulation mentioned above, but it is inconsistent with the violent merger model.

\section{Conclusions}\label{conclusion}
In this paper, we presented extensive photometric observations of SN 2014J collected from numerous ground-based telescopes as well as $HST$. The resulting light curves range from $\sim-$9 days to $\sim$+900 days from the maximum light, representing one of the few SNe Ia having such late-time observations. Our result confirms that SN 2014J is similar to a normal SN Ia around the maximum light, while it is distinguished by prominent blue-band emission in the early nebular phase. This excess emission can be well explained by additional SN light scattered by a disk-like CS dust located at a distance of a few times $\sim$ 10$^{17}$ cm. The CS dust of such a geometry can be formed by faint recurrent novae systems, although some DD scenario channels, such as mass outflows in final evolution or magnetically driven winds from the disk around the WD-WD system, cannot definitely be ruled out.

From t$\sim$+300 to +500 days, the luminosity of SN 2014J shows a fast decline compared to the $^{56}$Ni decay. 
After examining the evolution of the late-time spectra, we suggest that this behavior can be attributed to the evolution of [Fe III] $\lambda$4700 and [Fe II] $\lambda$5200 emissions, instead of the positron escape. We further analyze the very late-time $HST$ images of SN 2014J and confirm the late-time flattening of the light curve seen around t$\sim$+500 days. We fit the pseudo-bolometric light curve using a combination of radioactive decay isotopes  $^{56}$Ni, $^{57}$Ni and $^{55}$Fe. The derived $^{56}$Ni mass is in agreement with previous works,  and the $^{57}$Ni/$^{56}$Ni ratio is consistent with that predicted by the 2D and 3D delayed-detonation model simulations. Combined with the additional evidence from the excess emission in the blue band in the early nebular phase, we argue that the delayed-detonation through the SD scenario is favoured for SN 2014J. However, we caution that without late-time infrared observations we do not know the real flux fraction of the optical bands, and the pseudo-bolometric luminosity curve possibly do not represent the real bolometric luminosity evolution. Future late-time photometry and spectroscopy observations, especially those redward the optical bands, will help to discriminate among different explosion mechanisms and progenitor systems of SNe Ia.



\acknowledgments
This work is supported by the National Natural Science Foundation of China (NSFC grants 11325313, 11633002, and 11761141001)), and the National Program on Key Research and Development Project (grant no. 2016YFA0400803). J.-J. Zhang is supported by the National Science Foundation of China (NSFC, grants 11403096, 11773067), the Youth Innovation Promotion Association of the CAS, the Western Light Youth Project, and the Key Research Program of the CAS (Grant NO. KJZD-EW-M06). T.-M. Zhang is supported by the NSFC (grants 11203034, 11633002). We acknowledge the support of the staff of the Xinglong 80cm telescope. This work was also partially supported by the Open Project Program of the Key Laboratory of Optical Astronomy, National Astronomical Observatories, Chinese Academy of Sciences. Funding for the LJT has been provided by Chinese Academy of Sciences and the People's Government of Yunnan Province. The LJT is jointly operated and administrated by Yunnan Observatories and Center for Astronomical Mega-Science, CAS.

This paper is based on observations collected with the 80-cm Joan Or{\'o} Telescope (TJO) of the Montsec Astronomical Observatory (OAdM), which is owned by the Catalan Government and operated by the Institute for Space Studies of Catalonia (IEEC).

This work is also based on observations collected at Copernico and Schmidt telescopes (Asiago, Italy) of the INAF - Osservatorio Astronomico di Padova.
This paper is also partially based on observations collected at Copernico 1.82m telescope (Asiago, Italy) of the INAF-Osservatorio Astronomico di Padova; Galileo 1.22m telescope of the University of Padova; 
SB and LT are partially supported by the PRIN-INAF 2016 with the project ``Toward the SKA and CTA era: discovery, localisation, and physics of transient sources".
Partially based on observations collected at Copernico 1.82m telescope (Asiago, Italy) of the INAF - Osservatorio Astronomico di Padova.

N.E.R. acknowledges support from the Spanish MICINN grant ESP2017-82674-R and FEDER funds.

J.I.  acknowledges support from the Spanish MINECO grant ESP2017-82674-R.

SB, LT, PO are partially supported by the PRIN-INAF 2017 ``Towards the SKA and CTA era: discovery, localizazion, and physics of transient sorces" (PI Giroletti).

L. Wang is supported by NSF grant AST-1817099.

F. Huang was supported by the NSFC (11803021), and the Collaborating Research Program (OP201702) of the Key Laboratory of the Structure and Evolution of Celestial Objects, Chinese Academy of Sciences. 

The research of Y. Yang is supported through a Benoziyo Prize Postdoctoral Fellowship. 

We are grateful to Georgios Dimitriadis for kindly sharing us the spectrum of SN 2011fe.
\software{SWARP\ \ \citep{2002ASPC..281..228B},\ \ SExtractor\ \ \citep{1996A&AS..117..393B},\ \ SCAMP\ \ \citep{2006ASPC..351..112B},\ \ Matplotlib\ \ \citep{hunter},\ \ NumPy\ \ \citep{numpy},\ \ SciPy\ \ \citep{scipy}.}
{}

\startlongtable
\begin{deluxetable}{cccccccc}
\tablecolumns{6} \tablewidth{0pc} \tabletypesize{\scriptsize}
\tablecaption{Photometric Standards in SN 2014J Field \label{std} }
\tablehead{\colhead{Num.} &\colhead{$\alpha$(J2000)} &
\colhead{ $\delta$(J2000)} & \colhead{\textit{U} (mag)}&\colhead{\textit{B} (mag)} & \colhead{\textit{V} (mag)} & \colhead{\textit{R} (mag)} & \colhead{\textit{I} (mag)}} \startdata	
1&148.8956 & 69.6487 & 10.785(0.003) & 10.640(0.002) & 10.077(0.001) & 9.742(0.002) &  9.447(0.001)\\
2&149.1370 & 69.6547 & 12.808(0.005) & 12.849(0.004) & 12.303(0.002) & 11.953(0.004) & 11.621(0.003)\\
3&148.7262 & 69.6155 & 13.676(0.009) & 13.783(0.006) & 13.306(0.003) & 12.994(0.006) & 12.693(0.005)\\
4&148.8436 & 69.6332 & 14.231(0.014) & 14.326(0.009) & 13.816(0.004) & 13.468(0.007) & 13.122(0.006)\\
5&148.7226 & 69.6583 & 15.204(0.024) & 14.930(0.012) & 14.213(0.006) & 13.795(0.010) & 13.412(0.008)\\
6&148.9691 & 69.7350 & 15.717(0.038) & 15.536(0.018) & 14.850(0.008) & 14.421(0.014) & 14.005(0.012)\\
7&148.9235 & 69.6663 & 16.066(0.063) & 15.977(0.036) & 15.232(0.017) & 14.745(0.032) & 14.326(0.027)\\
8&148.8290 & 69.7257 & 17.625(0.186) & 17.262(0.072) & 16.499(0.030) & 15.986(0.052) & 15.603(0.044)\\
9&148.8551 & 69.6893 & 17.800(0.207) & 17.374(0.078) & 16.506(0.030) & 15.989(0.052) & 15.524(0.044)\\
10&148.8188 & 69.6409 & 16.174(0.069) & 16.367(0.042) & 16.141(0.021) & 16.037(0.043) & 15.876(0.035)\\
\enddata
\tablenotetext{}{Note: Uncertainties, in units of 0.001 mag, are $1\sigma$.}
\end{deluxetable}

\startlongtable
\begin{deluxetable}{cccccccc}
\tablecolumns{6} \tablewidth{0pc} \tabletypesize{\scriptsize}
\tablecaption{Ground-based Photometry of SN 2014J  \label{gphoto} }
\tablehead{\colhead{UT date} &\colhead{\tablenotemark{a}Epoch} &
\colhead{\textit{U} (mag)} & \colhead{\textit{B} (mag)} & \colhead{\textit{V} (mag)} & \colhead{\textit{R} (mag)}& \colhead{\textit{I} (mag)} & \colhead{Telescope} } \startdata
2014-01-22.9 & -9.7 & 13.31(04) & 12.88(02) & 11.63(03) & 10.91(06) & 10.49(05) & LJT\\
2014-01-23.9 & -8.7 & 13.01(04) & 12.62(02) & 11.39(03) & 10.62(06) & 10.19(03) & LJT\\
2014-01-24.8 & -7.8 & 12.79(04) & 12.43(02) & 11.21(02) & 10.48(06) & 10.08(03) & LJT\\
2014-01-25.1 & -7.5 & 12.72(03) & \nd & 11.13(03) & \nd & \nd & TJO\\
2014-01-25.9 & -6.7 & 12.69(04) & 12.29(02) & 11.07(01) & 10.40(06) & 9.96(03) & LJT\\
2014-01-25.9 & -6.7 & 12.70(02) & 12.33(04) & 11.11(01) & 10.47(01) & 10.04(03) & COP\\
2014-01-26.0 & -6.6 & 12.72(04) & 12.22(02) & 11.12(01) & 10.40(02) & 9.95(02) & TNG\\
2014-01-26.9 & -5.7 & 12.56(04) & 12.16(02) & 10.93(01) & 10.30(05) & 9.88(03) & LJT\\
2014-01-27.7 & -4.9 & \nd & 12.14(02) & 10.84(01) & 10.23(04) & 9.85(03) & TNT\\
2014-01-27.9 & -4.7 & 12.53(05) & 12.08(02) & 10.86(01) & 10.24(05) & 9.80(03) & LJT\\
2014-01-28.6 & -4.0 & \nd & 12.04(02) & 10.75(01) & 10.16(04) & 9.81(02) & TNT\\
2014-01-28.9 & -3.7 & 12.47(04) & 12.02(02) & 10.78(01) & 10.19(05) & 9.86(03) & LJT\\
2014-01-29.9 & -2.7 & 12.41(04) & 11.95(02) & 10.71(01) & 10.13(05) & 9.77(03) & LJT\\
2014-01-30.9 & -1.7 & 12.41(04) & 11.92(02) & 10.65(01) & 10.12(05) & 9.75(03) & LJT\\
2014-01-31.9 & -0.7 & 12.39(04) & 11.91(02) & 10.62(02) & 10.10(05) & 9.79(03) & LJT\\
2014-02-01.9 & +0.3 & 12.45(04) & 11.89(02) & 10.61(02) & 10.07(05) & 9.79(02) & LJT\\
2014-02-02.8 & +1.2 & 12.48(04) & 11.89(02) & 10.58(01) & 10.08(05) & 9.80(02) & LJT\\
2014-02-04.9 & +3.3 & 12.58(04) & 11.99(02) & 10.63(01) & 10.11(05) & 9.88(02) & LJT\\
2014-02-09.1 & +7.5 & 12.87(07) & 12.27(02) & 10.75(03) & 10.25(02) & 10.14(02) & TJO\\
2014-02-09.7 & +8.1 & \nd & 12.37(04) & 10.72(03) & 10.30(04) & 10.17(03) & TNT\\
2014-02-11.2 & +9.6 & \nd & 12.41(02) & 10.80(03) & 10.38(02) & 10.23(02) & TJO\\
2014-02-12.1 & +10.5 & 13.17(07) & 12.48(03) & 10.90(03) & 10.46(03) & 10.32(01) & TJO\\
2014-02-12.6 & +11.0 & 13.24(07) & 12.59(03) & 10.92(02) & 10.51(04) & 10.29(02) & TNT\\
2014-02-14.7 & +13.1 & \nd & 12.70(05) & 10.97(04) & 10.75(05) & 10.35(03) & TNT\\
2014-02-15.5 & +13.9 & 13.60(07) & 12.84(04) & 11.08(02) & 10.73(04) & 10.44(03) & TNT\\
2014-02-17.0 & +15.4 & 13.63(05) & 13.00(02) & 11.09(02) & 10.76(02) & 10.43(01) & TJO\\
2014-02-20.0 & +18.4 & 14.12(04) & 13.29(02) & 11.23(03) & 10.80(02) & 10.40(03) & TJO\\
2014-02-22.1 & +20.5 & 14.33(07) & 13.51(05) & 11.32(03) & 10.83(01) & 10.33(03) & TJO\\
2014-02-23.1 & +21.5 & 14.45(08) & 13.60(04) & 11.36(03) & 10.85(01) & 10.31(03) & TJO\\
2014-02-24.1 & +22.5 & 14.54(05) & 13.69(04) & 11.38(02) & 10.84(02) & 10.31(01) & TJO\\
2014-02-27.9 & +26.3 & 14.75(05) & 14.04(03) & \nd & \nd & \nd & TJO\\
2014-02-28.6 & +27.0 & 14.86(06) & 14.02(03) & 11.56(02) & 10.83(04) & 10.15(03) & TNT\\
2014-03-01.6 & +28.0 & 15.06(05) & 14.14(03) & 11.63(02) & 10.89(04) & 10.19(03) & TNT\\
2014-03-02.5 & +28.9 & 15.06(06) & 14.22(03) & 11.66(03) & 10.90(05) & \nd & TNT\\
2014-03-04.8 & +31.2 & 15.17(05) & 14.34(03) & 11.79(02) & 11.02(05) & 10.23(03) & TNT\\
2014-03-05.0 & +32.4 & \nd & 14.47(03) & 11.87(01) & 11.13(03) & 10.33(03) & TJO\\
2014-03-05.8 & +32.2 & \nd & 14.46(03) & 11.87(02) & 11.09(05) & 10.27(03) & TNT\\
2014-03-07.8 & +34.2 & 15.23(03) & 14.57(03) & 11.99(03) & 11.29(04) & 10.46(04) & TJO\\
2014-03-09.8 & +36.2 & 15.41(04) & 14.65(03) & 12.15(03) & 11.41(02) & 10.60(01) & TJO\\
2014-03-10.8 & +37.2 & 15.40(04) & 14.75(03) & 12.18(02) & 11.48(02) & 10.66(01) & TJO\\
2014-03-11.1 & +37.5 & 15.35(20) & 14.71(03) & 12.29(03) & 11.46(04) & 10.56(03) & COP\\
2014-03-12.0 & +38.4 & 15.57(02) & 14.77(02) & 12.22(03) & 11.54(02) & 10.72(02) & TJO\\
2014-03-13.0 & +39.4 & 15.56(05) & 14.79(02) & 12.26(02) & 11.58(03) & 10.82(01) & TJO\\
2014-03-13.6 & +40.0 & \nd & 14.83(03) & 12.28(03) & 11.55(05) & 10.81(03) & TNT\\
2014-03-13.8 & +40.2 & \nd & 14.84(02) & 12.31(03) & 11.62(02) & 10.87(02) & TJO\\
2014-03-14.6 & +41.0 & 15.66(08) & 14.84(03) & 12.33(02) & 11.62(05) & 10.82(03) & TNT\\
2014-03-15.5 & +41.9 & 15.76(09) & 14.87(04) & 12.38(02) & 11.66(05) & 10.86(03) & TNT\\
2014-03-19.5 & +45.9 & 15.81(06) & 14.90(03) & 12.51(02) & 11.82(05) & 11.09(03) & TNT\\
2014-03-20.5 & +46.9 & 15.86(06) & 14.96(03) & 12.52(02) & 11.84(05) & 11.13(03) & TNT\\
2014-03-22.5 & +48.9 & 15.83(06) & 14.95(03) & 12.60(02) & 11.92(05) & 11.22(03) & TNT\\
2014-03-23.0 & +49.4 & 15.84(04) & 14.98(02) & 12.59(03) & 11.95(02) & 11.30(01) & TJO\\
2014-03-25.5 & +51.9 & 15.82(08) & 14.96(03) & 12.67(02) & 12.02(05) & 11.35(03) & TNT\\
2014-03-26.5 & +52.9 & 15.85(11) & \nd & 12.71(02) & 12.06(05) & 11.40(03) & TNT\\
2014-03-27.9 & +54.3 & 15.90(04) & \nd & 12.78(03) & 12.12(02) & 11.54(02) & TJO\\
2014-03-28.6 & +55.0 & 15.91(19) & 15.00(03) & 12.77(02) & 12.13(05) & 11.49(03) & TNT\\
2014-03-29.5 & +55.9 & 15.98(07) & 15.06(03) & 12.77(02) & 12.14(05) & 11.51(03) & TNT\\
2014-03-30.5 & +56.9 & 16.08(07) & 15.05(03) & 12.83(02) & 12.17(05) & 11.58(03) & TNT\\
2014-04-01.0 & +58.4 & 16.07(13) & 15.04(12) & 12.98(04) & 12.15(04) & 11.77(10) & COP\\
2014-04-05.1 & +62.5 & 16.00(04) & 15.07(03) & 12.93(03) & 12.32(02) & 11.81(02) & TJO\\
2014-04-07.9 & +65.3 & \nd & 15.06(12) & 13.16(02) & 12.40(04) & 12.10(02) & COP\\
2014-04-08.9 & +66.3 & 16.25(04) & \nd & 13.10(03) & 12.50(03) & 12.09(02) & TJO\\
2014-04-09.5 & +66.9 & \nd & \nd & 13.12(02) & 12.48(05) & 11.98(03) & TNT\\
2014-04-11.9 & +69.3 & 16.20(06) & \nd & 13.14(03) & 12.58(03) & 12.17(02) & TJO\\
2014-04-13.5 & +70.9 & \nd & 15.15(05) & 13.18(02) & 12.59(05) & 12.12(03) & TNT\\
2014-04-17.9 & +75.3 & 16.26(05) & 15.27(03) & 13.31(03) & 12.79(02) & 12.36(01) & TJO\\
2014-04-19.5 & +76.9 & 16.55(10) & 15.25(03) & 13.36(02) & 12.79(05) & 12.34(03) & TNT\\
2014-04-20.6 & +78.0 & \nd & 15.29(03) & 13.42(02) & 12.84(05) & 12.44(03) & TNT\\
2014-04-21.6 & +79.0 & 16.55(09) & 15.30(03) & 13.42(02) & 12.85(05) & 12.42(03) & TNT\\
2014-04-22.5 & +79.9 & 16.54(09) & 15.30(03) & 13.46(02) & 12.88(05) & 12.46(03) & TNT\\
2014-04-24.6 & +82.0 & 16.66(16) & 15.29(03) & 13.47(02) & 12.95(05) & 12.52(03) & TNT\\
2014-04-26.6 & +84.0 & \nd & 15.32(03) & 13.56(02) & 12.99(05) & 12.57(03) & TNT\\
2014-04-27.6 & +85.0 & \nd & 15.37(03) & 13.58(02) & 13.03(05) & 12.68(03) & TNT\\
2014-04-28.6 & +86.0 & 16.74(07) & 15.40(03) & 13.61(02) & 13.07(05) & 12.66(03) & TNT\\
2014-05-01.9 & +89.3 & \nd & 15.45(16) & 13.60(06) & 13.10(07) & 12.72(13) & TJO\\
2014-05-13.6 & +101.0 & \nd & 15.54(04) & 13.95(02) & 13.49(05) & 13.13(03) & TNT\\
2014-05-14.6 & +102.0 & \nd & 15.61(04) & 14.03(02) & 13.52(05) & 13.15(03) & TNT\\
2014-05-26.6 & +114.0 & 17.55(10) & 15.79(02) & 14.28(01) & 13.82(05) & 13.47(03) & TNT\\
2014-06-08.5 & +126.9 & 17.73(18) & \nd & 14.56(02) & 14.08(05) & 13.88(04) & TNT\\
2014-06-10.9 & +129.3 & \nd & 15.81(10) & 14.49(03) & \nd & \nd & COP\\
2014-09-18.2 & +228.7 & \nd & 17.59(03) & 16.67(04) & 16.67(06) & 16.02(07) & NOT\\
2014-10-27.0 & +268.4 & \nd & 17.95(03) & 17.23(02) & \nd & \nd & COP\\
2014-10-30.8 & +271.2 & \nd & 18.05(06) & 17.18(06) & 17.18(10) & 16.18(12) & LJT\\
2014-11-21.8 & +293.2 & \nd & 18.62(07) & 17.27(12) & 17.21(18) & 16.22(14) & LJT\\
2014-12-18.9 & +320.3 & \nd & 18.70(05) & 17.96(08) & 18.06(13) & 16.70(14) & LJT\\
2014-12-20.1 & +321.5 & \nd & 18.83(05) & 17.83(03) & \nd & \nd & COP\\
2015-01-19.8 & +352.2 & \nd & 19.05(03) & 18.44(04) & \nd & \nd & COP\\
2015-01-22.9 & +355.3 & \nd & 19.19(06) & 18.39(04) & 18.31(08) & 16.98(10) & LJT\\
2015-03-10.8 & +402.2 & \nd & 19.91(16) & 18.92(10) & \nd & \nd & COP\\
2015-03-31.9 & +423.3 & \nd & \nd & 19.06(10) & \nd & \nd & COP\\
2015-04-11.8 & +434.2 & \nd & \nd & 19.10(14) & \nd & \nd & COP\\
\enddata
\tablenotetext{a}{Days after $t_{Bmax}$ on 2014-02-02.1 (JD 2456690.1).}
\end{deluxetable}

\startlongtable
\begin{deluxetable}{cccc}
\tablecolumns{6} \tablewidth{12pc} \tabletypesize{\scriptsize}
\tablecaption{Log of observations of SN 2014J with \textit{HST} WFC3/UVIS \label{hst} }
\tablehead{\colhead{Filter} &
\colhead{\qquad Date of Obs.\qquad\qquad} & \colhead{Exp. Time (s)} & \colhead{\tablenotemark{a}Epoch} } \startdata
F438W& 2014-09-05 19:12:57 & 2$\times$256 & 216.2  \\
F555W& 2014-09-05 19:29:44 & 2$\times$256 & 216.2  \\
F555W& 2015-02-02 05:06:06 & 3$\times$128 & 365.6 \\
F438W& 2015-02-02 05:24:41 & 3$\times$512 & 365.6 \\
F555W& 2015-07-20 01:35:40 & 3$\times$144 & 533.5 \\
F438W& 2015-07-20 01:55:15 & 3$\times$448 & 533.5 \\
F555W& 2016-01-02 05:29:17 & 3$\times$144 & 699.6 \\
F438W& 2016-01-02 06:42:27 & 3$\times$448 & 699.7 \\
F555W& 2016-07-02 01:37:07 & 1720         & 881.5 \\
F438W& 2016-07-02 04:07:56 & 2420         & 881.6 \\
\enddata
\tablenotetext{a}{Days after $t_{Bmax}$ on 2014-02-02.1 (JD 2456690.1).}
\end{deluxetable}

\startlongtable
\begin{deluxetable}{ccccc}
\tablecolumns{4} \tablewidth{0pc} \tabletypesize{\scriptsize}
\tablecaption{\textit{HST} Photometry of SN 2014J (AB magnitude) \label{hphoto} }
\tablehead{ \colhead{Epoch} &\colhead{\textit{F438W} (mag)} & \colhead{\textit{F555W} (mag)} & \colhead{$\Delta$mag/100 days of $F438W$}  & \colhead{$\Delta$mag/100 days of $F555W$}} \startdata
 216.2&17.476(0.001) & 16.331(0.001) & - & - \\
 365.6&19.672(0.001) & 18.668(0.002) & 1.47(0.01) &1.57(0.01)\\
 533.5&22.223(0.012) & 21.330(0.008) & 1.52(0.01) &1.43(0.01)\\
 699.6&24.017(0.031) & 22.454(0.023) & 1.08(0.03) &0.68(0.02)\\
 881.6&24.988(0.044) & 23.513(0.023) & 0.53(0.05) &0.58(0.03)\\
\enddata
\end{deluxetable}

\begin{table}[htbp]
\caption{Decline rates measured during 50 $-$ 110 days after the $B$-band maximum light of SN 2014J\label{decline}}
\centering
\begin{tabular*}{18cm}{@{\extracolsep{\fill}}cccccc}
\hline\hline 
Band & $U$ &$B$ &$V$ &$R$ &$I$ \\
\hline
Decline Rate (50-110d) in mag/100d & 2.56$\pm$0.08  & 1.16$\pm$0.05 & 2.50$\pm$0.04 & 2.86$\pm$0.03 & 3.39$\pm$0.06\\
\hline
\end{tabular*}
\end{table}

\begin{table}[htbp]
\caption{Best-fit Parameters of equation (\ref{gaussian})  \label{fit} }
\centering
\begin{tabular*}{9cm}{@{\extracolsep{\fill}}ccc}
\hline\hline 
$A$ & $\tau$(days) &$\sigma$(days)  \\
\hline
$0.12 \pm 0.03$ & $64 \pm 8$ & $ 26 \pm 8$ \\
\hline
\end{tabular*}
\end{table}

\begin{table}[htbp]
\caption{Parameters of the Dust Structure\label{parameter} }
\centering
\begin{tabular*}{10cm}{@{\extracolsep{\fill}}cc}
\hline\hline 
Parameter & Range  \\
\hline
$\theta_{obs}$   &  [10$^{\circ}$, 60$^{\circ}$]    \\      
$\theta_{disk}$   &  [15$^{\circ}$, 30$^{\circ}$]\\
$R_{in}$/light day  &  [20, 110] \\
$R_{wid}$/light day &  [20, 110]\\
$\tau_B$ &  [0.2, 2.0]\\
\hline
\end{tabular*}
\end{table}
 
\begin{table}[htbp]
\caption{Results from Monte Carlo Simulation\label{fit result2} }
\centering
\begin{tabular*}{10cm}{@{\extracolsep{\fill}}ccccc}
\hline\hline 
$\theta_{obs}$ & $\theta_{disk}$ & $R_{in}$/light day &  $R_{wid}$/light day & $\tau_B$ \\
\hline
\multicolumn{5}{c}{Disk 1}                            \\ 
30.0$^{\circ}$   &  15$^{\circ}$ & 40 & 40 & 0.9 \\
60.0$^{\circ}$   &  15$^{\circ}$ & 40 & 40 & 0.9 \\

\hline
\multicolumn{5}{c}{Disk 2}                            \\ 
30.0$^{\circ}$   &  30$^{\circ}$ & 110 & 100 & 0.6 \\
60.0$^{\circ}$   &  30$^{\circ}$ & 110 & 100 & 0.6 \\
\hline
\end{tabular*}
\end{table}

\begin{table}[htbp]
\caption{Late-time Decline Rate of the Bolometric Light Curve of SN 2014J\label{decline rate} }
\centering
\begin{tabular*}{10cm}{@{\extracolsep{\fill}}cc}
\hline\hline 
Period(days) & decline rate($\Delta$mag/100 days)  \\
\hline
277 $-$ 365   &  1.37 $\pm$ 0.02    \\      
365 $-$ 416   &  1.60 $\pm$ 0.03\\
416 $-$ 533   &  1.43 $\pm$ 0.02\\
533 $-$ 649 &  1.08 $\pm$ 0.04\\
649 $-$ 700 &  0.84 $\pm$ 0.07\\
700 $-$ 796 &  0.64 $\pm$ 0.04\\
796 $-$ 881   &  0.45 $\pm$ 0.03\\
881 $-$ 983   &  0.40 $\pm$ 0.05\\
\hline
\end{tabular*}
\end{table}

\clearpage

\appendix
\begin{figure}[htbp]
\center
\includegraphics[width=.5\linewidth]{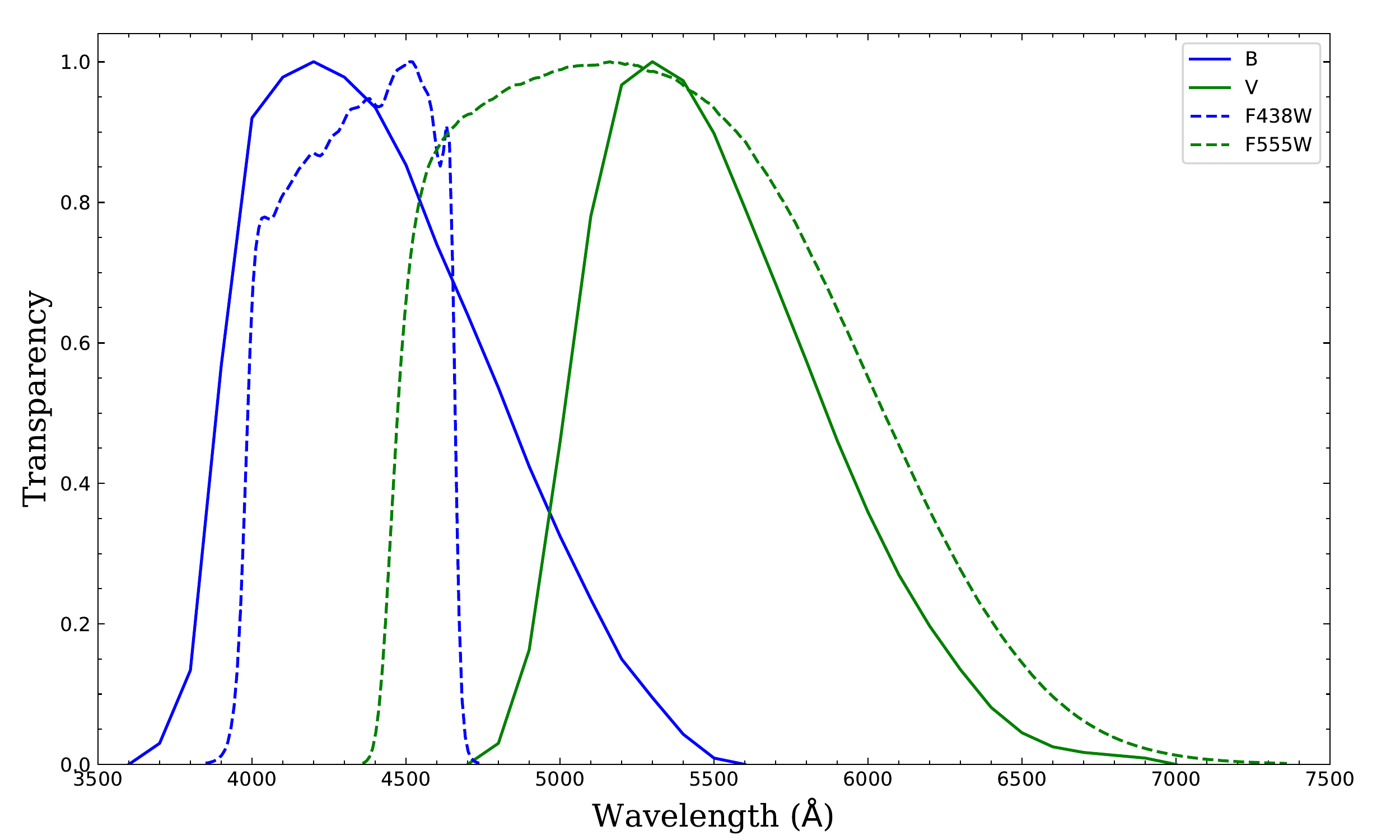}
\vspace{0.2cm}
\caption{The transmission curves of Johnson $B$, $V$, F438W and F555W bands. All the curves are normalized to the peak.}
\label{BV_H} \vspace{-0.0cm}
\end{figure}

\end{document}